%% file: proceedings.tex
\documentclass[sigconf, authorversion, nonacm]{acmart}

\usepackage{multirow}
\usepackage{enumitem}
\usepackage{xspace}
\usepackage[nolist]{acronym}
\usepackage{placeins}
\usepackage[FIGTOPCAP]{subfigure}

\usepackage[acronym]{glossaries}
\usepackage{colortbl}
\glsdisablehyper

\newcommand{\fref}[1]{Figure~\ref{#1}}
\newcommand{\tref}[1]{Table~\ref{#1}}
\newcommand{\afref}[1]{Supplementary Figure~\ref{#1}}
\newcommand{\atref}[1]{Supplementary Table~\ref{#1}}
\newcommand{\acrexpl}[1]{\acs{#1}: \acl{#1}}

\newcommand{\EARLY}[0]{\textsc{Early Warning}\xspace}
\newcommand{\ABOVE}[0]{\textsc{Above Limit}\xspace}

\newcommand{\meanstd}[2]{#1\,$\pm$\,#2}
\newcommand{\gdL}[1]{#1\,g/dL}
\newcommand{\mgL}[1]{#1\,mg/L}

\newcommand{\US}{US\xspace}
\newcommand{\EU}{EU\xspace}

\definecolor{early}{HTML}{a5c8e1}
\definecolor{above}{HTML}{aca3cd}
\definecolor{margin}{HTML}{fdb678}

\begin{acronym}
\acro{BAC}[BAC]{blood alcohol concentration}
\acro{BrAC}[BrAC]{breath alcohol concentration}
\acro{alcolock}[alcolock]{alcohol ignition interlock device}
\acro{CAN}{Controller Area Network}
\acro{ML}{machine learning}
\acro{ROC}{receiver operating characteristics}
\acro{PRC}{precision-recall curve}
\acro{AUC}{area under the curve}
\acro{AUROC}{area under the receiver operating characteristic curve}
\acro{PERCLOS}{Percentage Eye Closure}
\acro{SHAP}{SHapley Additive exPlanations}
\acro{LOSO}{leave-one-subject-out}
\acro{IIJA}{Infrastructure Investment and Jobs Act}
\acro{HCI}{Human–Computer Interaction}

\acro{AUPRC}{area under the precision-recall curve}
\acro{CAN-bus}{controller area network bus}
\acro{Euro NCAP}{European New Car Assessment Programme}
\acro{AUDIT}{Alcohol Use Disorder Identification Test}
\acro{PEth}{phosphatidylethanol}
\acro{GSR}{General Safety Regulation}
\acro{M}{mean}
\acro{SD}{standard deviation}
\acro{CI}{confidence interval}
\acro{SVM}{support vector machine}
\acro{NHTSA}{National Highway Traffic Safety Administration}
\acro{WHO}{World Health Organization}
\acro{EU}{European Union}
\acro{DADDS}{Driver Alcohol Detection System for Safety}
\acro{IQR}{interquantile range}
\acro{s}{second}
\acro{RMS}{root mean square}
\acro{FP}{false positive}
\acro{FN}{false negative}
\acro{TP}{true positive}
\acro{TN}{true negative}
\acro{E}{Eye movement}
\acro{H}{Head movement}
\acro{MLP}{multi-layer perceptron}
\acro{NOALCOHOL}[{no alcohol}]{\emph{no alcohol} (\ac{BAC} of \gdL{$0.00$})}
\acro{MODERATE}[{moderate}]{\emph{moderate}, that is, below the \ac{WHO} recommended legal limit (i.e., \gdL{$0.00$} $<$~\ac{BAC} $\leq$~\gdL{$0.03$})}
\acro{SEVERE}[{severe}]{\emph{severe}, that is, above the \ac{WHO} recommended legal limit (i.e., \gdL{$0.05$} $<$~\ac{BAC} $\leq$~\gdL{$0.07$})}
\acro{ABS}[ABS]{anti-lock braking system}
\acro{ESP}[ESP]{Electronic Stability Control}
\end{acronym}


\AtBeginDocument{%
  \providecommand\BibTeX{{%
    \normalfont B\kern-0.5em{\scshape i\kern-0.25em b}\kern-0.8em\TeX}}}

\setcopyright{None}

\begin{document}

\title[Preprint - Machine learning using driver monitoring cameras to detect drunk driving]{Preprint - Leveraging driver vehicle and environment interaction: Machine learning using driver monitoring cameras to detect drunk driving}


\author{Kevin Koch}
\email{kevin.koch@unisg.ch}
\authornote{Joint first author.}
\orcid{0000-0003-4523-2668}
\affiliation{%
\institution{University of St. Gallen}
\streetaddress{Dufourstrasse 40a}
\city{St. Gallen}
\country{Switzerland}
\postcode{9000}
}

\author{Martin Maritsch}
\authornotemark[1]
\orcid{0000-0001-9920-0587}
\email{mmaritsch@ethz.ch}
\affiliation{%
\institution{ETH Zürich}
\streetaddress{Weinbergstrasse 56/58}
\city{Zürich}
\country{Switzerland}
\postcode{8092}
}

\author{Eva van~Weenen}
\orcid{0000-0001-5500-2108}
\email{evanweenen@ethz.ch}
\affiliation{%
\institution{ETH Zürich}
\streetaddress{Weinbergstrasse 56/58}
\city{Zürich}
\country{Switzerland}
\postcode{8092}
}

\author{Stefan Feuerriegel}
\orcid{0000-0001-7856-8729}
\email{feuerriegel@lmu.de}
\affiliation{%
\institution{LMU Munich}
\streetaddress{Geschwister-Scholl-Platz 1}
\city{Munich}
\country{Germany}
\postcode{80539}
}

\author{Matthias Pfäffli}
\orcid{0000-0003-2712-8672}
\email{matthias.pfaeffli@irm.unibe.ch}
\affiliation{%
\institution{University of Bern}
\streetaddress{Murtenstrasse 26}
\city{Bern}
\country{Switzerland}
\postcode{3008}
}

\author{Elgar Fleisch}
\orcid{0000-0002-4842-1117}
\email{efleisch@ethz.ch}
\affiliation{%
\institution{ETH Zürich and University of St. Gallen}
\streetaddress{Weinbergstrasse 56/58 and Dufourstrasse 40a}
\city{Zürich and St. Gallen}
\country{Switzerland}
\postcode{8092 and 9000}
}

\author{Wolfgang Weinmann}
\orcid{0000-0001-8659-1304}
\authornote{Joint last author.}
\email{wolfgang.weinmann@irm.unibe.ch}
\affiliation{%
\institution{University of Bern}
\streetaddress{Murtenstrasse 26}
\city{Bern}
\country{Switzerland}
\postcode{3008}
}

\author{Felix Wortmann}
\orcid{0000-0001-5034-2023}
\authornotemark[2]
\authornote{Corresponding author.}
\email{felix.wortmann@unisg.ch}
\affiliation{%
\institution{University of St. Gallen}
\streetaddress{Dufourstrasse 40a}
\city{St. Gallen}
\postcode{9000}
\country{Switzerland}
}

\renewcommand{\shortauthors}{Koch and Maritsch, et al.}

\begin{abstract}
Excessive alcohol consumption causes disability and death. Digital interventions are promising means to promote behavioral change and thus prevent alcohol-related harm, especially in critical moments such as driving. This requires real-time information on a person’s \acl{BAC} (\acs{BAC}). Here, we develop an in-vehicle machine learning system to predict critical \acs{BAC} levels. Our system leverages driver monitoring cameras mandated in numerous countries worldwide. We evaluate our system with $n=30$ participants in an interventional simulator study. Our system reliably detects driving under any alcohol influence (\acl{AUROC} [\acs{AUROC}] $0.88$) and driving above the \acs{WHO} recommended limit of \gdL{$0.05$} \acs{BAC} (\acs{AUROC} $0.79$). Model inspection reveals reliance on pathophysiological effects associated with alcohol consumption. To our knowledge, we are the first to rigorously evaluate the use of driver monitoring cameras for detecting drunk driving. Our results highlight the potential of driver monitoring cameras and enable next-generation drunk driver interaction preventing alcohol-related harm.
\end{abstract}



\maketitle

\input{body/00_introduction}
\input{body/01_related}
\input{body/02_data}
\input{body/03_ml}
\input{body/04_results}
\input{body/05_discussion}
\input{body/06_conclusion}

\subsection*{Data availability}
The following procedure is required by our local ethics committee. Any requests for raw data (i.e., blood alcohol concentrations, camera data, driving data, de-identified patient characteristics) will be reviewed by the scientific study board leading the involved research group. Only applications for non-commercial use will be considered and should be sent to the corresponding author. Applications should outline the purpose for the data transfer. Any data that can be shared will need approval from the scientific study board and a Material Transfer Agreement in place. All data shared will be de-identified.

\begin{acks}
This research was funded and supported by the Bosch IoT Lab at the University of St.~Gallen and ETH Zurich. The views and conclusions contained in this document are those of the authors and should not be interpreted as necessarily representing the official policies, either expressed or implied, of the sponsors. M. Maritsch acknowledges funding from the Swiss National Science Foundation (183569). E.v. Weenen acknowledges funding from the Swiss National Science Foundation (186932). The funding bodies had no control over design, conduct, data, analysis, review, reporting, or interpretation of the research conducted.
\end{acks}

\bibliographystyle{ACM-Reference-Format}
\bibliography{proceedings}

\clearpage

\input{body/98_appendix.tex}

\end{document}

%% file: body/00_introduction.tex
\section{Introduction}
Alcohol consumption is responsible for 5\% of the global disease burden and is further the cause of 1 in 20 deaths worldwide~\cite{whoalcohol2019}. To promote behavior change, digital interventions provide effective means to prevent harm in critical situations due to alcohol consumption and intoxication~\cite{npjalcohol2021, trackingalcohol2018, shani2015, shani2017}. In particular, digital interventions could promote behavior change by delivering real-time targeted feedback on alcohol consumption. However, to intervene early, real-time predictions of alcohol consumption are needed. 

Alcohol consumption increases, among others, the risk of traffic crashes, making drunk driving one of the leading causes of severe crashes on public roads. For example, in the \US, around 30 people die each day in traffic crashes in which one of the parties is under the influence of alcohol, and, together, alcohol-related crashes amount to 30\% of all traffic fatalities~\cite{ustraffic2020}. To prevent alcohol-related crashes, in-vehicle systems are needed to detect drunk driving and enable targeted interventions. Examples of such interventions are, e.g., warnings of impairment and forced stops of the vehicle.

As of today, the only reliable measurement technology for identifying intoxicated driving are ignition interlock devices that analyze a driver's breath alcohol. However, ignition interlock devices are expensive and, further, require regular maintenance. State-of-the-art devices cost around USD 1000 and need yearly maintenance~\cite{radun2014, swov2005}. Regulators are aware of these challenges: For example, the \US Congress recently introduced a concrete timeline for drunk driving detection systems in vehicles~\cite{infrastructurebill2021}, calling for scalable, low-cost, and easily accessible technologies.  

A cost-effective and scalable approach could be to measure the driver performance on the basis of the existing sensor technology of today's vehicles. Even though progress has been made toward fully autonomous driving, experts agree that autonomous driving will not be widely available in the next two decades~\cite{autonomousadaption2017, autonomous2018}. Hence, in the coming years, driving will still require to interact with the vehicle as well as the environment, and, thus, detection systems are needed that build upon existing vehicle technology. Here, we develop and evaluate a machine learning system for detecting drunk driving based on driver monitoring cameras already built into modern vehicles. In fact, driver monitoring cameras will be introduced in the coming years in almost all new vehicles due to safety regulations, such as the \ac{Euro NCAP} or the \EU \ac{GSR}, which make them mandatory from 2024 onwards~\cite{encaproadmap2017, gsr2019}.

\subsection*{Contributions}

In this paper, we develop and evaluate a novel machine learning system to detect drunk driving from driver monitoring cameras leveraging driver vehicle and environment interaction. Specifically, our system extracts information on gaze behavior and head movements from driver monitoring cameras and then predicts whether drivers exceed two critical thresholds for \ac{BAC}: (1)~\ac{BAC} values above \gdL{$0.00$} yet below the \ac{WHO} recommended~\cite{whosafer2019} legal limit of \gdL{$0.05$} for early warnings and (2)~\ac{BAC} values above the \ac{WHO} recommended limit. For this purpose, we conducted an interventional clinical trial with $n=30$ healthy participants that completed driving tasks in a research-grade driving simulator with different levels of alcohol intoxication and in different driving environments.

The contribution, novelty, and significance of our work are as follows:
\begin{itemize}
    \item \textbf{Contributions:} (1) Our system reliably detects driving under the influence of alcohol. (2) Our approach based on a driver monitoring camera outperforms previous approaches based on driving data (e.g., ~\cite{lee2010, drunk_driving_related_1, drunk_driving_related_2, drunk_driving_related_9, drunk_driving_related_5}) by a clear margin. (3) Our system is highly generalizable as the detection is robust to unseen individuals and driving scenarios. (4) Analysis of the learned patterns of our machine learning model shows that our system relies upon known pathophysiological mechanisms of alcohol intoxication.
    \item \textbf{Novelty:} To our knowledge, we are the first to rigorously evaluate the use of driver monitoring cameras for detecting drunk driving by conducting a clinical trial with participants driving in a research-grade simulator, both sober and drunk. Although existing work on driver state monitoring addresses safety-critical states such as drowsiness or distraction, previous attempts at detecting drunk driving with vehicle signals have not achieved sufficiently good results or were not rigorously evaluated (e.g.,~\cite{lee2010, not_evaluated2016, not_driving2022, not_evaluated2021, drunk_driving_not_eval2021, drunk_driving_related_1, drunk_driving_related_2, drunk_driving_related_3}).
    \item \textbf{Significance:} Our system offers a viable approach based on existing technologies, allowing for a rapid implementation to prevent potential negative consequences of drunk driving after decades of stagnating high alcohol-related road crashes without any significant advancement by regulators and industry~\cite{ustraffic2020, destatis2020}. To accelerate this development, we provide the source code of our evaluated machine learning system on GitHub: \sloppy\url{https://github.com/im-ethz/CHI-2023-paper-Leveraging-driver-vehicle-and-environment-interaction}.
\end{itemize}

%% file: body/01_related.tex
\begin{figure*}%
\centering
\subfigure[][]{%
\label{fig:example-1}%
\includegraphics[height=60mm]{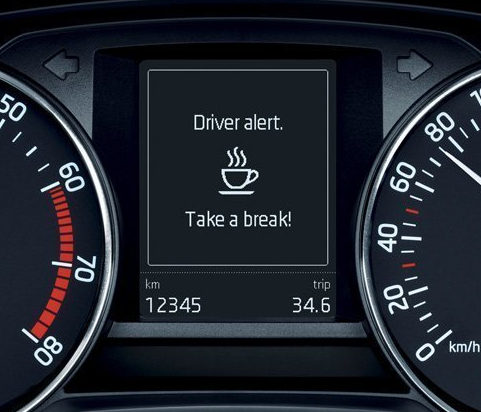}}%
\hspace{10pt}%
\subfigure[][]{%
\label{fig:example-2}%
\includegraphics[height=60mm]{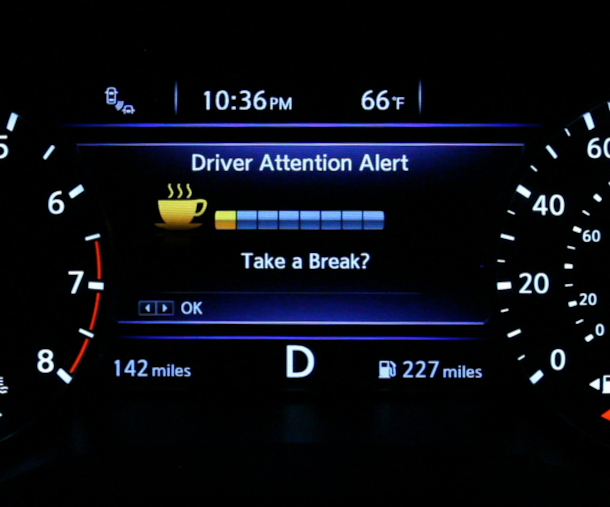}}%
\Description{Two examples of existing drowsiness and attention warning systems in today's cars. In Figure 1a a warning is issued from the control cluster. The warning consists of a coffee cup as an image which is surrounded by the text ``Driver alert. Take a break!''. In Figure 1b, a warning is issued from the control cluster. The warning consists of a coffee cup with a vertical bar indicating the level of attention which is surrounded by the text ``Driver Attention Alert. Take a Break?''.}
\caption[]{\textbf{Examples of existing drowsiness and attention warning systems in today's cars.} \subref{fig:example-1} Drowsiness warning delivered in Volkswagen cars; \subref{fig:example-2} Attention warning delivered in Nissan cars.}
\label{fig:examples}%
\end{figure*}

\section{Related Work}
In recent years, the \ac{HCI} community has already taken steps toward digital systems for monitoring alcohol consumption and intervening when needed. Previous work detects alcohol intoxication based on gait analysis~\cite{gaitanalysis2012}, smartphone interactions and related contextual data~\cite{smartphoneinteractiongames2020, smartphonecontext2021, drunk2018}, sometimes in conjunction with smart breathalyzer systems~\cite{connectedbreathalyzer2017, connectedbreathalyzer2018}, wrist-worn devices that measure physical or transdermal activity~\cite{transdermalwrist2019, wrist2020}, or social media data~\cite{social_media2019, social_media2018}. Interventions targeting responsible alcohol consumption include self-management of use through diaries~\cite{SoberDiary2015}, informing peers and family about recent drinking behavior~\cite{social2018, social2019}, or virtual agents that intervene~\cite{virtualagent2013}. These systems demonstrate that it is feasible to detect alcohol consumption and that digital interventions are capable of successfully averting harmful behavior. However, the use scenarios of these systems are very broad; we believe that addressing harmful alcohol use in situations associated with truly harmful outcomes will significantly impact reducing alcohol-related harm.

\subsection{Overview on driver state detection}

In the realm of driving, research and industry have introduced increasingly advanced assistance and safety systems in recent decades that understand the interaction between drivers, vehicles, and environments. In the case of driver monitoring systems, there has been a strong focus on the condition of drowsiness. For several years, there are now systems commercially available that issue drowsiness warnings based on existing vehicle signals from the CAN bus (i.e., a central communication network with high-frequency messages between vehicle components, such as steering wheel or accelerator/brake pedal actuation, speed, or acceleration)~\cite{bosch2022, drowsinesssystems2020}. Due to increasing availability and higher detection performance, modern commercial drowsiness detection systems shifted to the use of cameras~\cite{smarteye2022, drowsinesssystems2020}. Building on these widespread driver monitoring systems, research and industry have developed, validated, and commercialized various algorithms that cover additional use cases, often coming from non-vehicle domains, such as driver identification, distraction detection, and emotion detection based on eye, pupil or head movements, or facial expressions~\cite{driveridentification2018, distraction2016, affectiva2022, driverbehaviorcameras2022}. Common across these systems is their warning approach when an impairment is detected. They intervene with a simple audiovisual warning, delivered via the vehicle's infotainment system on the center console, which emits a clearly audible warning tone and recommends taking a break. \fref{fig:examples} shows the implementation of such warnings by Volkswagen and Nissan~\cite{vw2022, nissan2022}.

\subsection{Camera-based driver monitoring}
Driver monitoring systems in industrial research are often proprietary approaches, and implementation details are not available to a broad audience. Fortunately, related academic work provides extensive research on driver monitoring (e.g., see the literature reviews in ~\cite{driverbehaviorcameras2022, drowsinesssystems2020}). Generally, camera-based driver monitoring focuses on a per-use-case solution of problems, and, if solved, a specific use case is combined with existing systems as part of an ensemble to monitor various driver states~\cite{bosch_dms2022, smarteye2022}. In the following, we discuss the specific example of drowsiness detection to explain common approaches on the development and evaluation of driver monitoring systems. In general, the majority of detection systems rely on the detection of behavioral patterns that are associated with a specific driver state. In the case of drowsiness, these patterns are, for example, higher percentages of eye closure, yawning, or head nodding and scaling~\cite{driverbehaviorcameras2022}. Hence, related studies recorded people acting and imitating typical drowsiness patterns using (driver monitoring) cameras (e.g., openly available datasets are~\cite{yaw_data2014, drowsiness_data2016}). Subsequently, machine learning approaches are trained to detect these patterns. In general, two major approaches exist. First, traditional approaches rely upon basis signals describing low-level characteristics of visual behavior, such as gaze direction, eyelid closure, or facial landmarks, which are extracted with computer vision algorithms (e.g., OpenCV~\cite{opencv} and Viola Jones~\cite{viola_jones2001}) or neural networks (e.g.,~\cite{gaze_deep2018, deep_face2016}). Subsequently, those signals are used to either hand-craft features, such as \ac{PERCLOS}, or to train deep neural networks for the automated detection of drowsiness patterns (e.g.,~\cite{drowsiness_example2020, deep_learning_drowsiness2019}). Second, more recent work leverages end-to-end neural networks with the intention of directly detecting behavior patterns associated with drowsiness from raw images (e.g.,~\cite{raw_deeplearning_drowsiness2018}). If specific movements exceed predefined thresholds, the system raises an alarm.

While the prior described imitation of drowsy patterns by people acting has safety advantages and can be conducted with less effort (i.e., drivers do not drive in a critical drowsiness state), these imitations are only approximations. In comparison, statistical analyses on drowsiness detection commonly rely upon driving in actual drowsiness by either using long driving times or methods for sleep deprivation (e.g., see literature review~\cite{statistics_drowsiness2020}). To our knowledge, there are only very few studies on drowsiness detection based on cameras in which drivers were actually in a drowsy state~\cite{drowsiness_state2019, drowsiness_similar_pipeline2019, lee2014}. Here, the entire visual, facial, and head movement behavior is used as a proxy for driver impairment. Awake and drowsy driving trips are partitioned into smaller sequences over which behavior signals are aggregated.

\begin{table*}
\Description{The table shows different information about the participants. Overall, $n=30$ participants (15 female, 15 male) aged \meanstd{37.03}{9.24} years participated. People had a body weight of \meanstd{75.00}{26.0}\,kg and size of \meanstd{173.5}{56.4}\,cm. They had a driving experience of \meanstd{13.3}{7.8}~years and drove \meanstd{5.71}{2.57}~kilometers per year. One participant was a professional truck driver. Participant had \acf{AUDIT} scores of \meanstd{5.71}{2.57} and none were above the \acf{PEth} limit of 180\,ng/mL.}
	\caption{\textbf{Participant characteristics ($n=30$).} \acrexpl{AUDIT}; \acrexpl{PEth}; \acrexpl{SD}.}~\label{sup:tab:participant-characteristics}%
 \label{tab:participants}
 \centering
	\begin{tabular}{l|l|l|l}
	\toprule
	& \textbf{\meanstd{Mean}{\ac{SD}}} & \textbf{Min} & \textbf{Max}\\
	\hline
    Gender & 15 female, 15 male & --&--\\
    Age [years] & \meanstd{37.0}{9.2} & 21 & 59\\
    Weight [kg] & \meanstd{75.0}{26.0} & 55.7 & 96.3 \\
    Height [cm] & \meanstd{173.5}{56.4} & 159 & 192 \\
    Driving experience [years] & \meanstd{13.3}{7.8} & 4 & 29\\
    Driving distance [km/year] & \meanstd{8300}{6820} & 400 & 30000\\
    Professional drivers & 1 (truck driver) & --&--\\
    \acs{AUDIT} score  & \meanstd{5.71}{2.57} & 1 & 11\\
    \acs{PEth} blood concentration [ng/mL] & 4 below detection and 5 below quantification limit\textsuperscript{*} & --&-- \\
    & 21 above quantification limit\textsuperscript{*} with \meanstd{84.1}{48.7} & 28.6 & 199.0 \\
    \bottomrule
    \multicolumn{4}{l}{}\\
	\multicolumn{4}{l}{\footnotesize{\textsuperscript{*}detection limit 10\,ng/mL, quantification limit: 20\,ng/mL}}
	\end{tabular}
\end{table*}

\subsection{Drunk driving detection}
Although there is considerable work on drunk driving, this work focuses largely on statistical analysis examining the influence of alcohol on driving performance (see, for example, the systematic review in~\cite{jongen2016}). These empirical findings provide a basis for regulators to decide on relevant legal thresholds for driving under the influence of alcohol~\cite{alcohol_limit_005}. However, these previous works did not focus on the real-time identification of drunk driving, which would allow for intervening when a driver is actually drunk. Prior research also covered the negative effect of alcohol on gaze behavior while driving (e.g., tunnel vision)~\cite{moskowitz2000, shiferaw2019}.

In contrast, related work on \emph{detecting} drunk driving is still at comparatively early stages. Similar to the early work on detecting drowsiness, research focused on detecting drunk driving based on driving behavior such as steering, pedal usage, and vehicle speed. These past findings demonstrate the general feasibility of detecting drunk driving from in-vehicle signals~\cite{lee2010}. Unfortunately, the results of this study and comparable work on the detection of drunk driving~\cite{drunk_driving_related_1, drunk_driving_related_2, drunk_driving_related_3, drunk_driving_related_4, drunk_driving_related_5, drunk_driving_related_6, drunk_driving_related_7, drunk_driving_related_8, drunk_driving_related_9} show that, while machine learning models trained on driving behavior yield a good performance on previously seen drivers, they do not achieve the performance needed to generalize to unseen drivers. One main reason is that driving behavior is overlaid by additional influences from the environment (highway, rural, and urban) and further varies across drivers (e.g., slow vs. fast drivers, defensive vs. aggressive drivers), which introduces additional noise and thus makes inferences of alcohol levels challenging. As a remedy, we propose to shift from driving behavior (i.e., how does the vehicle behave?) to driver behavior (i.e., how does the driver behave?). To this end, we leverage driver monitoring cameras to capture eye movements, gaze events, and head movements as fine-grained and frequent predictors of driving under the influence of alcohol.

Regarding camera-based driver monitoring to detect drunk driving, we found that related work either lacks rigor and clarity or only proposes a solution without evaluation. More specifically, studies miss key information about experimental designs, drinking procedures, targeted alcohol levels, or methods to measure alcohol levels. Furthermore, they apply unclear evaluation approaches, making it impossible to interpret the results, replicate these studies, or understand them at all (e.g.,~\cite{alcohol_unclear_study2018, little_data2020, drunk_driving_related_7, very_small_sample2022, not_driving2022}). Although many studies elaborate on the possibility of detecting drunk driving and design systems, they often do not evaluate them (e.g.,~\cite{drunk_driving_not_eval2021, not_evaluated2021, not_evaluated2016}).

Given the limitations of previous work on drunk driving detection, we conducted an interventional clinical trial following standardized principles in clinical research to rigorously evaluate driver monitoring cameras for drunk driving detection. In general, the standard regarding study quality in the field of driver monitoring cameras is insufficient as shown by both literature reviews on drowsiness and drunk driving detection. Ultimately, we believe it is necessary to analyze drivers in a reliable impairment state to thoroughly evaluate detection approaches.

%% file: body/02_data.tex
\section{Data collection}

In this paper, we aimed at developing and evaluating a novel \ac{ML} system to detect critical \ac{BAC} thresholds. For this purpose, we conducted a non-randomized, single-blinded, interventional, single-center study (\href{https://clinicaltrials.gov/ct2/show/NCT04980846}{ClinicalTrials.gov NCT04980846}) called DRIVE (design and implementation of a drunk driving detection system). The DRIVE study took place between August 2021 and November 2021 in Bern, Switzerland. The study followed the Declaration of Helsinki, the guidelines of good clinical practice, the Swiss health laws, and the ordinance on clinical research. Each participant gave informed written consent prior to any study-related procedure. The study was further approved by the local ethics committee in Bern, Switzerland (ID 2021-00759). Throughout this paper, we report statistics as \meanstd{mean}{\ac{SD}}. In the following, we describe the overall study procedure and data collection in detail.

\subsection{Sample size calculation}
To determine the number of participants, we cannot rely on power calculation as traditional null hypothesis testing is not applicable to our study (i.e., there is no null hypothesis for the development of \ac{ML} models). Therefore, we implemented an established methodology from a previous study~\cite{sample_size2012} to extrapolate the discriminatory power of \ac{ML} with increasing sample size. Due to the lack of pre-existing literature in the field, this method was applied to preliminary data that we retrieved in a pilot study ($n=10$). Based on this approach, an \ac{AUROC} of $0.85$ to detect driving above the \ac{WHO} recommended \ac{BAC} of \gdL{$0.05$} was projected for a sample size of $30$.

\begin{figure*}
\centering
    \includegraphics[width=0.85\textwidth]{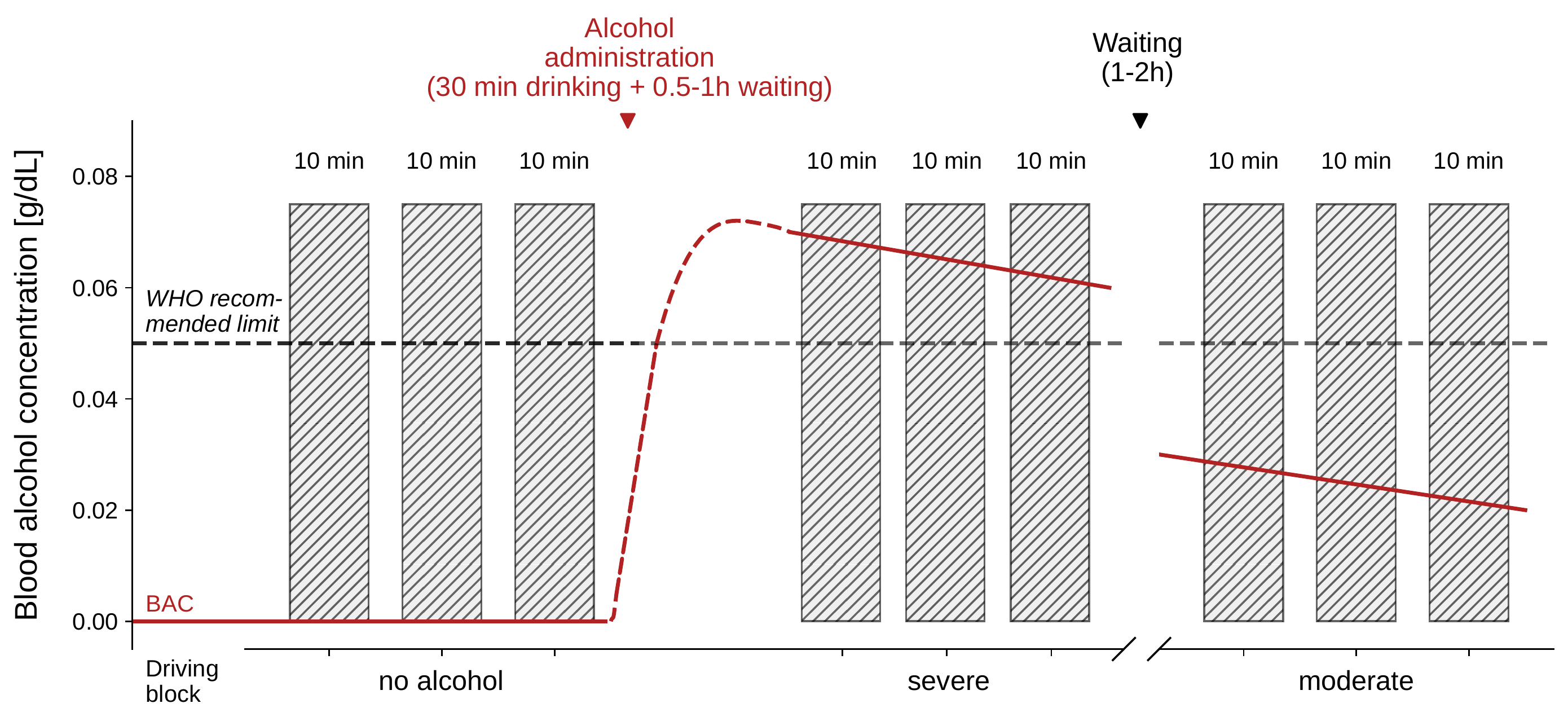}
\Description{Study overview showing the targeted blood alcohol levels in relation to the driving blocks. On the x-axis, the three driving blocks each with their three driving phases (driving scenarios) are presented. Each driving phase lasted 10 minutes, resulting in 30 minutes per driving block. Between the first driving block (no alcohol) and the second driving block (severe) took the alcohol administration place, which lasted 30 minutes for drinking and 30 minutes to 1 hour waiting time to reach the blood alcohol level. Between the second driving block (severe) and the third driving block a waiting time between 1 to 2 hours is indicated. On the y-axis the blood alcohol concentration between 0.00 and 0.08 g/dL. At a blood alcohol value of 0.05 g/dL is a vertical dashed line drawn, which indicates the WHO recommended limit. The first driving block should took place at a blood alcohol level of 0.0 g/dL, the second driving block at a blood alcohol level of 0.07 g/dL after having a peak at 0.08 g/dL blood alcohol level and the third driving block should then start at 0.03 g/dL blood alcohol level.}        
\caption{\textbf{Study overview.} Overview of study procedure where participants performed driving arranged in three blocks during which different alcohol levels were targeted (i.e., no alcohol, moderate, and severe). \acrexpl{BAC}; \acrexpl{WHO}.}
\label{fig:study_plan}
\end{figure*}

\subsection{Participants}
Inclusion criteria for individuals eligible for participation were: (1)~passing the driver examination at least 2 years before study inclusion; (2)~possession of a driver's license that is valid in the European Union or Switzerland; and (3)~reporting moderate alcohol consumption (i.e., neither total absence nor excess). The latter was validated based on two instruments. First, participants were asked with the \ac{AUDIT}~\cite{audit1993} multiple-choice questionnaire about their alcohol consumption (e.g., ``How often do you have a drink containing alcohol?'') and related experiences (e.g., ``During the past year, how often have you been unable to remember what happened the night before because you had been drinking?''). Each answer is scored between 0 to 4 and their total sum is a proxy for drinking behavior. A sum of 0 indicates total absence. A score of 1 to 7 suggests low-risk consumption. Scores from 8 to 14 suggest hazardous or harmful alcohol consumption and a score of 15 or more indicates the likelihood of alcohol dependence (excess). Second, we collected the \ac{PEth} level of a capillary blood sample on the pre-screening day, which needed to be below 210\,ng/mL~\cite{wurst2015}. Exclusion criteria for participation in the study were if participants met one or more of the following: pregnancy or an intention for pregnancy; health conditions incompatible with alcohol consumption; known or suspected non-compliance; participation in another study with investigational drug preceding and during the present study; personal dependencies with the study team (e.g., employees, family members, and other dependent persons); experience of motion sickness.

In total, we screened 39 participants for eligibility in our study. After checking for inclusion and exclusion criteria, we finally collected data from $n=30$ participants (15 female, 15 male, age \meanstd{37.03}{9.24} years). Detailed participant characteristics are reported in \tref{tab:participants}, summarizing the demographics, driving experience, and alcohol consumption of the participants. More details on the enrollment in the study are in Appendix~\ref{app:details_participants}.

\begin{figure*}
\centering
    \includegraphics[width=0.85\textwidth]{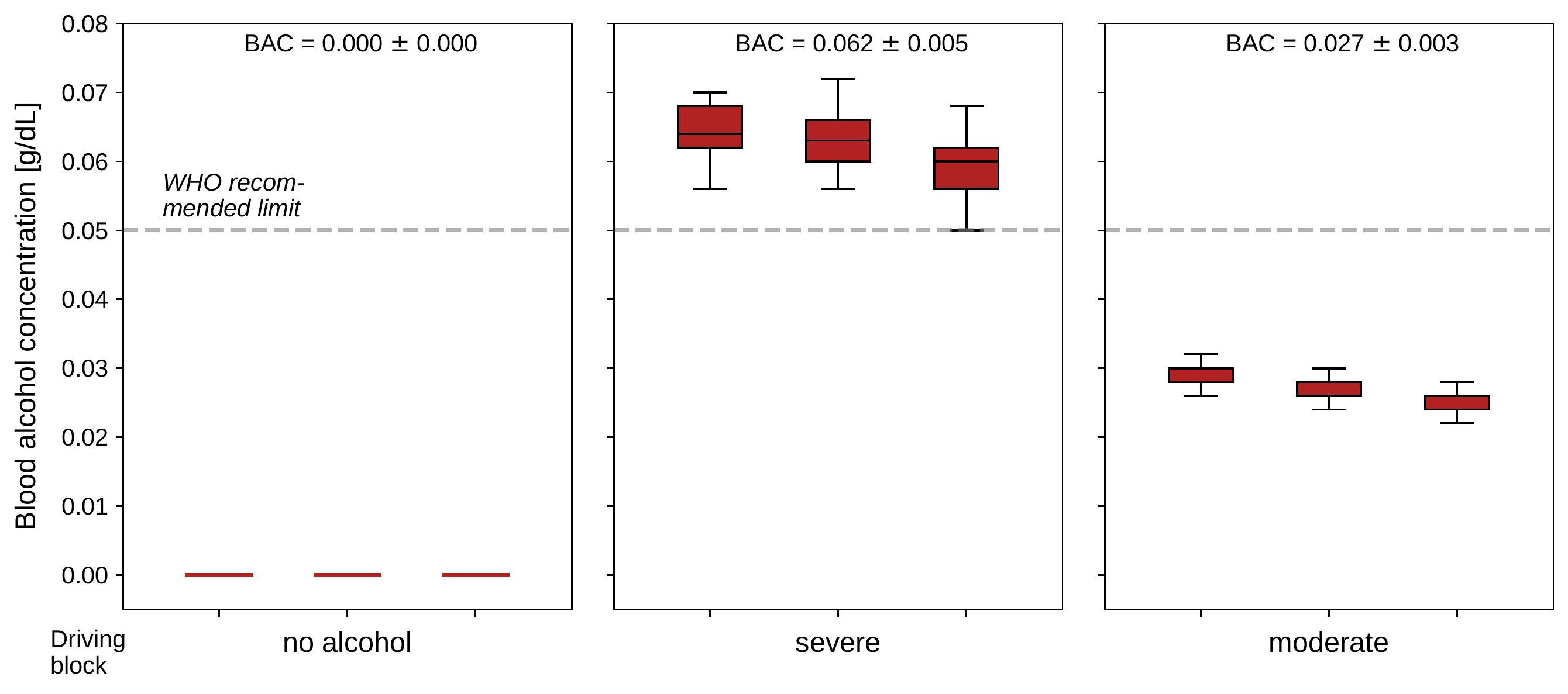}
    \label{fig:second}
\Description{Actual blood alcohol levels presented as a boxplot chart during the three driving blocks. On the x-axis, the three driving blocks each with their three driving phases (driving scenarios) are presented. On the y-axis the blood alcohol concentration between 0.00 and 0.08 g/dL. At a blood alcohol value of 0.05 g/dL is a vertical dashed line drawn, which indicates the WHO recommended limit. In the first driving block (no alcohol), all drivers had in each driving phase actual blood alcohol levels of 0.00 g/dL. In the second driving block (severe), the drivers had across the driving phases actual blood alcohol levels between 0.07 g/dL and 0.05 g/dL. In the third driving block (moderate), the drivers had actual blood alcohol levels between 0.03 g/dL and 0.02 g/dL.}
\caption{\textbf{Blood alcohol levels.} Observed \acs{BAC} levels across participants during driving. \acrexpl{BAC}; \acrexpl{WHO}.}
\label{fig:bac_levels}
\end{figure*}

\subsection{Study procedure}
Following a telephone screening interview, participants were invited to the study location for on-site screening. After informed consent was obtained, a simulator training session followed. The training session was used to familiarize participants with the driving simulator and to test whether they experience motion sickness.

On the day of the study visit, participants arrived at the research facility in the morning after an overnight fast. After an initial, additional training session in the driving simulator, participants conducted driving tasks in a driving simulator during controlled alcohol administration. Here, the targeted alcohol levels were: (1)~\ac{NOALCOHOL}; (2)~\ac{SEVERE}; and (3)~\ac{MODERATE}. No driving took place between \ac{BAC} levels of \gdL{0.03} and \gdL{0.05}. For the \ac{SEVERE} condition, we defined that participants should be above the \ac{WHO} recommended \ac{BAC} legal limit of \gdL{$0.05$}, since this (or a lower) limit is mandated by 97 countries worldwide~\cite{whoalcohol2019}. While higher \ac{BAC} levels would likely increase the expected effects on driving behavior, it would put participant's health at unnecessary risk~\cite{moskowitz2000}. For the \ac{MODERATE} condition, we relied on past research indicating altered driving behavior already at a \ac{BAC} of \gdL{$0.02$}~\cite{moskowitz2000}.

For each alcohol level, the participants drove for 30~minutes in a research-grade driving simulator. The driving was split across three scenarios (highway, rural, and urban). Driving in each scenario lasted for 10~minutes, separated by breaks of 1--2~minutes for intermediate breath alcohol measurements. During driving, participants were captured by a driver monitoring camera. Participants further had breaks of 1--2 hours between each driving block to reduce the potential effects of drowsiness. Participants received food and non-caffeinated drinks during the study day. The procedure for alcohol administration and driving sessions is shown in \fref{fig:study_plan}.

\subsection{Alcohol administration and measurement}
To obtain specific target levels for \ac{BAC}, we used established procedures for alcohol administration~\cite{lee2010}. Depending on the gender, weight, and age of a participant, we calculated the amount of alcohol to be administered with an updated version of the established Widmark formula~\cite{widmark1932, watson1981, brick2006}. Afterward, participants received the calculated amount of alcohol divided over three mixed drinks (vodka and a non-alcoholic beverage in equal parts) that all had to be consumed within 30 minutes at a steady pace. To achieve comparable conditions for all participants, that is, the same target \ac{BAC} of \gdL{$0.07$} when the driving starts, we administered the amount of alcohol for a \ac{BAC} of \gdL{$0.08$} and then waited until the \ac{BAC} dropped to our target of \gdL{$0.07$}. The additional amount further ensured that physiological differences (e.g., resorption deficits~\cite{resorptiondeficit2000}) were mitigated. Participants were informed before the study that they would consume alcohol during the study, but they were blinded to the total amount and their current \ac{BAC} values during the procedure.

Alcohol levels of the participants were measured throughout the study with a certified and calibrated breath alcohol measurement device (Dräger Alcotest 6820, Drägerwerk AG \& Co KGaA, Lübeck, Germany), that is authorized to be used by law enforcement in Switzerland~\cite{atemalkoholschweiz2016}. We measured the \ac{BAC} of participants prior to the first driving session (\emph{no alcohol}) and started to measure again 20 minutes after the intake of the last alcoholic beverage (i.e., before the second driving session) to avoid that mouth alcohol can influence the \ac{BrAC} measurement~\cite{20minutesbreath}. Measurements were conducted repeatedly (every 2--5 minutes) until two consecutive measurements were at the target \ac{BAC} of \gdL{$0.07$} or below. To guarantee that the \ac{BAC} is in the target range for each driving block, participants conducted \ac{BAC} measurements prior to each driving scenario. After the second driving block, participants had their first food consumption during a long break in which the study team occasionally measured the \ac{BAC} (every 2--30 minutes) depending on how close participants were to the target level for the third and last driving session (\ac{BAC} of \gdL{$0.03$}). As soon as two consecutive measurements were at or below a \ac{BAC} of \gdL{$0.03$}, participants commenced the last driving session.

In each of the three driving blocks (i.e., no alcohol, severe, moderate), participants drove for a total of 30 minutes. The time was split across three scenarios (highway, rural, and urban), each with a duration of 10 minutes and in random order. Before each scenario, the \ac{BrAC} was measured, which thus resulted in three measurements per driving block. To convert between \ac{BrAC} and \ac{BAC}, we used a factor of $0.2$ (e.g., a \ac{BrAC} of \mgL{$0.35$} corresponds to a \ac{BAC} of \gdL{$0.07$}) as defined by national law in Switzerland~\cite{conversionschweiz2012}. The corresponding observations of \ac{BAC} are reported in \fref{fig:bac_levels}. In our clinical study, we recorded a \ac{BAC} of \gdL{\meanstd{$0.000$}{$0.000$}} during the first driving block (\ac{NOALCOHOL}); \gdL{\meanstd{$0.062$}{$0.005$}} during the second driving block (\ac{SEVERE}); and \gdL{\meanstd{$0.027$}{$0.003$}} during the third driving block (\ac{MODERATE}). Hence, the observed \ac{BAC} levels are within the intended target ranges.

\subsection{Driving simulator}

\begin{figure}%
\centering
\includegraphics[height=54mm]{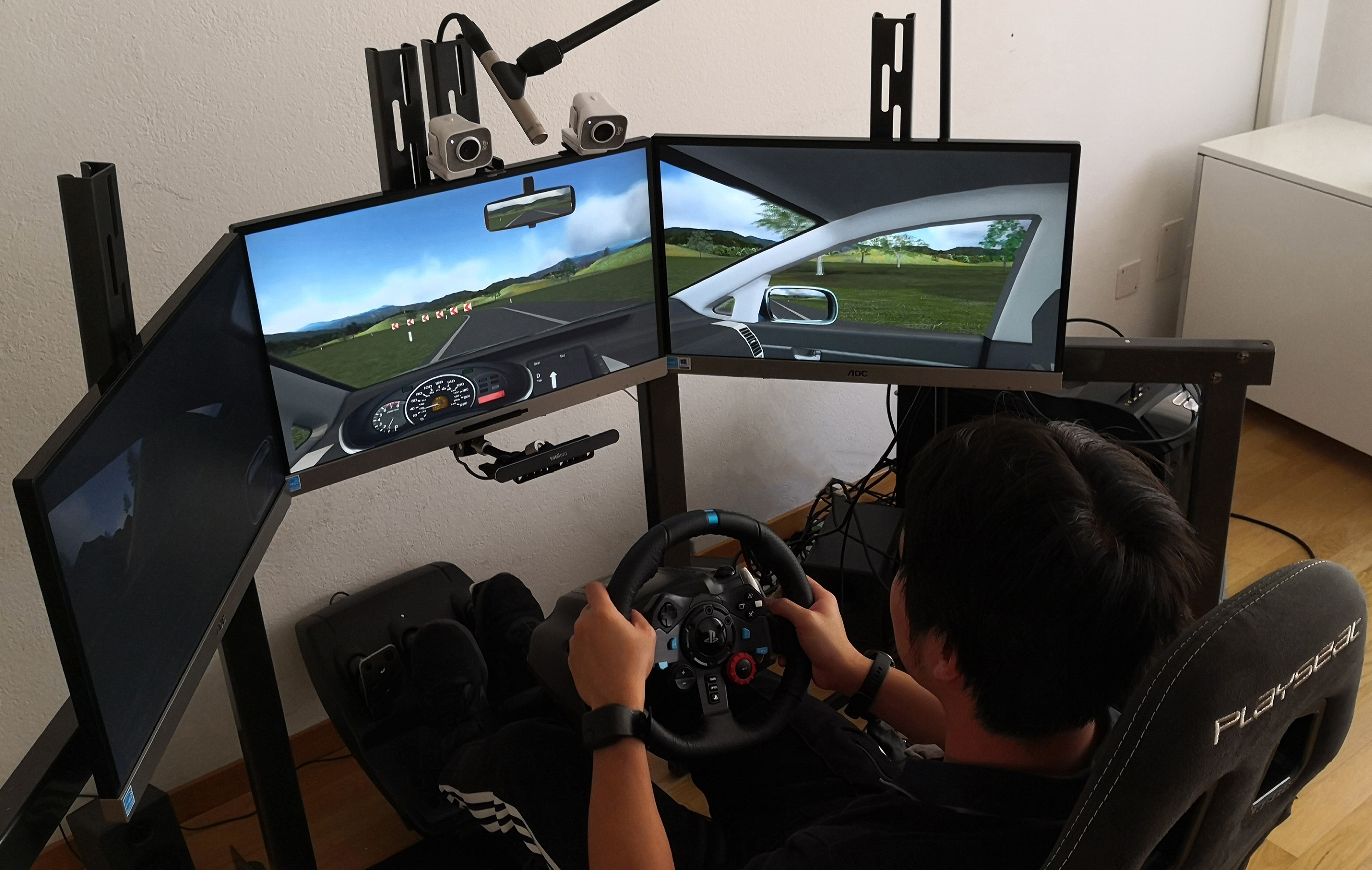}%
\Description{The simulator consists of three screens with a field-of-view of 270 degrees, a steering wheel, pedals, and a driver seat. Below the center screen the driver monitoring camera is mounted.}
\caption{\textbf{Driving simulator setup.} Driving simulator with the driver monitoring camera mounted below the center screen.}
\label{fig:simulator}%
\end{figure}

The driving tasks were conducted in a driving simulator (Carnetsoft BV, Groningen, The Netherlands). The simulator offers a realistic setting and is widely used in medical research~\cite{winsum2019, diabetes_driving_simulator}. The simulator consists of three screens with a field-of-view of 270 degrees, a steering wheel, and pedals for an immersive driver experience. All participants used automatic transmission. The driver's field-of-view mimics that of a real vehicle, including a dashboard with all standard instruments as well as rear-view mirrors. The driving simulator setup is shown in \fref{fig:simulator}.

To cover a variety of driving situations, we implemented three different driving scenarios with distinct characteristics: {highway}, {rural}, and {urban}. These are as follows: (1)~The highway scenario comprised of a two-lane highway with one-way traffic. Here, the route was mostly straight with a few wide curves. The speed limit varied between 80 and 120\,km/h. Drivers experience varying traffic densities, ranging from free flow to slow-moving traffic. (2)~The rural scenario consisted of two-lane rural roads with traffic in both directions and several intersections with and without yield signs. The speed limit was between 60 and 100\,km/h. Drivers experienced other traffic participants and had to react to occasional events, such as a stopping bus or slower speeds in front of a school. (3)~The urban scenario was used to reflect driving in a city. The route consisted of shorter and narrower roads compared to the two other scenarios. In addition, there were a large number of warning signs, intersections with and without yield or stop signs, and special events, such as pedestrians crossing streets. The speed limit was between 30 and 50\,km/h. In all scenarios, variations of the following parameters were randomly assigned within limits: traffic density, behavior of other traffic participants, traffic light circuits, maneuvers of other road users, vehicle types, and traffic at intersections. The order of the driving scenarios was randomized for each driver in every driving block. The routes in each driving scenario were intentionally kept the same across all three blocks. To avoid learning effects over time during the experiment itself, people had an extensive training session before the experimental procedure to make them familiar with the routes. In line with existing research, we kept the routes in each driving scenario the same because of two fundamental reasons: (1)~familiarity with routes is often an excuse for drunk driving~\cite{decision_drunk_driving_1993}, and (2)~known routes are often associated with an increased chance of alcohol-related road crashes~\cite{close_home2017}.

Participants were instructed to adhere to local traffic laws (in Switzerland), act as they would in normal road traffic, and make use of all provided vehicle facilities, e.g., turn signal lights. While driving, participants had to follow the guidance of an on-board navigation system.

\subsection{Collecting gaze behavior and head movements}
During driving, gaze behavior and head movements were recorded using a driver monitoring camera. Here, we used an infrared camera system (Tobii Pro Nano, Tobii AB, Stockholm, Sweden), which was directly mounted below the center screen of the driving simulator. The camera comes with a pre-validated eye tracking algorithm that calculates the gaze positions of drivers as Cartesian coordinates on the center screen with a frequency of 60\,Hz. Moreover, this algorithm for eye tracking calculates the current position of the eyes, which we used to infer head movements. At the beginning of each driving day, the eye tracker was calibrated for the participant with respect to the center screen of the driving simulator.

%% file: body/03_ml.tex
\section{Machine learning system}
\begin{figure*}
\centering
    \includegraphics[width=\textwidth]{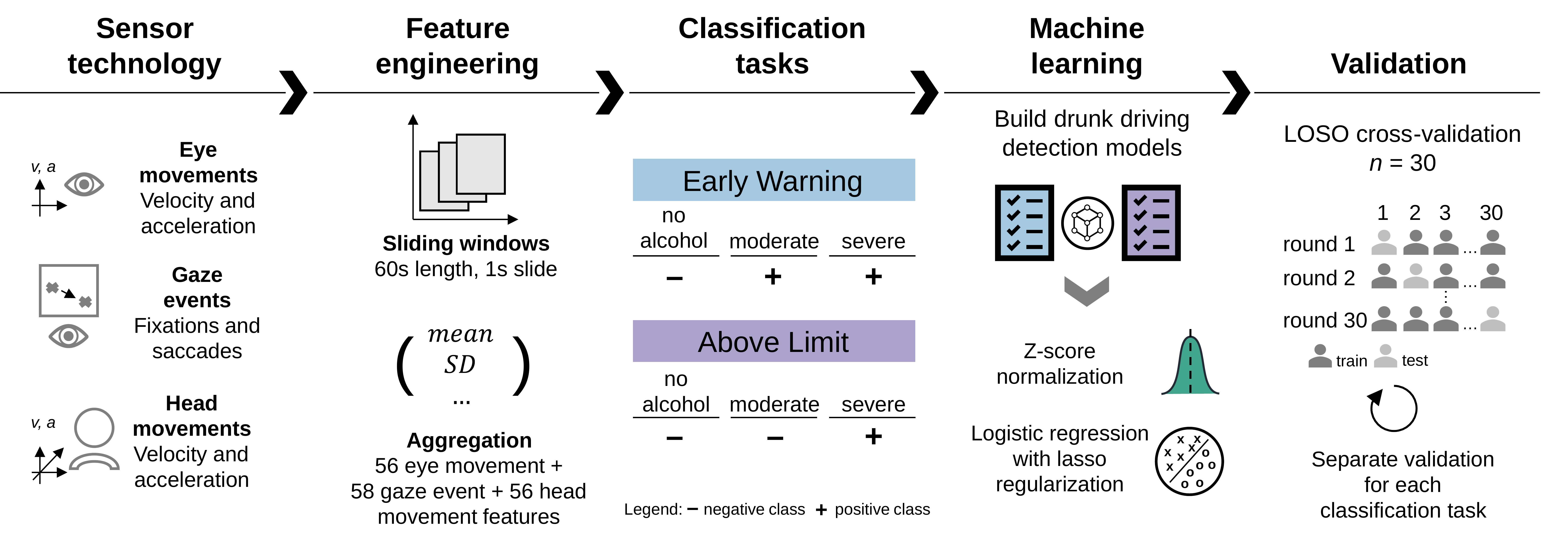} 
\Description{Overview over the machine learning system. The system and its validation consists of five different steps. In the first step, we describe the sensor technology and what kind of features were calculated (eye movements, gaze events, and head movements). In the second step, the sliding window aggregation based on 60 seconds window length in 1 second steps. For each sliding window, different feature aggregation functions, such as mean or standard deviation, are used. In the third step, the classification tasks EARLY WARNING and ABOVE LIMIT are shown. In the fourth step, the model is explained. For each classification tasks, separate models are built. These models use z-score normalization and logistic regression with lasso regularization. In the fifth step, the leave-one-subject-out cross-validation is explained. It shows that for each classification task thirty models are built always leaving a different user out for validation.}
\caption{\textbf{Machine learning system and evaluation.} Overview of our machine learning system based on driver monitoring cameras to detect drunk driving at different thresholds. \acrexpl{LOSO}; \acrexpl{SD}.}
\label{fig:pipeline}
\end{figure*}

We developed a novel machine learning system to detect drunk driving using driver monitoring cameras (see \fref{fig:pipeline}). Our system proceeds as follows: In the first step, the driver monitoring camera is used to capture information on gaze behavior (velocity, acceleration, fixations, saccades) and head movements. Second, feature engineering is applied using a sliding window approach to train a machine learning model. For training, we varied the underlying classification task (i.e, the label of the prediction), so that two different \ac{BAC} thresholds are classified as ``drunk driving''. In the following, we explain the single steps of this system in detail.

\subsection{Feature generation}
We intentionally chose intelligible features that allow for post~hoc explainability in order to interpret the \ac{ML} model against previous knowledge about pathophysiological mechanisms. In our system, we use three feature groups, which are able to reflect well-known pathophysiological changes due to an alcohol intoxication~\cite{moskowitz2000, fixation2020}. The feature groups are: (1)~eye movements, (2)~gaze events, and (3)~head movements. To compute the final features, we first pre-processed the gaze behavior and head movement data into high-frequent signals, which we then summarized into feature vectors for our \ac{ML} model using sliding windows and statistical aggregation functions. This procedure is explained in the following.  

\subsubsection{Data pre-processing}
We first calculated the following signals from the gaze behavior and the head movements: (1)~\underline{Eye movements}. We calculated the velocity and acceleration of vertical, horizontal, and the combination of both gaze directions. In addition, we used the absolute vertical and horizontal coordinate positions on the screen. (2)~\underline{Gaze events}. We distinguish fixations and saccades. Fixations are time periods in which drivers concentrate their gaze on a certain point or region, whereas saccades are rapid movements of a driver's gaze after and before fixations. To identify both, we applied the REMoDNaV algorithm~\cite{remodnav2021}. The REMoDNaV algorithm calculates the duration for each fixation and saccade. Further, it identifies the amplitude (i.e., distance traveled) as well as the peak and average velocity for each of these gaze events. (3)~\underline{Head movements}. We computed the velocity and acceleration of the head across three dimensions: vertical, horizontal, and depth (i.e., distance to the eye tracker). We further aggregated them into a combined vector. 

\subsubsection{Sliding window and feature calculation}
We used a sliding window approach to split the time-series data into time windows~\cite{slidingwindow}. The window length is subject to an inherent trade-off. On the one hand, long time windows capture more variance and should thus be more informative. On the other hand, short time windows are necessary for near real-time predictions and thus early warnings. Informed by prior literature on detecting driver drowsiness~\cite{drowsiness_similar_pipeline2019}, we set the window length to 60 seconds with a shift of 1 second. We perform a sensitivity analysis with other window lengths in \fref{fig:sensitivity}d) and Appendix~\ref{app:robustness-checks}, providing empirical evidence supporting our choice.  

Each window was then processed by different aggregation functions in order to map the time-series data onto single features. Here, we used the following mathematical functions (see Appendix~\ref{app:feature-generation}): mean, standard deviation, 0.05 and 0.95 quantiles, skewness, kurtosis, and power (i.e., sum of squares divided by the amount of each signal or event). We intentionally preferred the 0.05 and 0.95 quantiles over minimum and maximum, respectively, as we found the former to be more robust to outliers. For gaze event features, we additionally counted the overall number of fixations and saccades. As a result, we have 56 features for eye movements, 58 features for gaze events, and 56 features for head movements. In total, this approach led to 151,200 samples (30 subjects $\times$ 81 minutes driving [i.e., 90 minutes of driving, but in the first 60 seconds of each driving phase no window was created] $\times$ every second a window aggregating 60 seconds).

\subsection{Predictive modeling}

For \ac{ML}, we make use of two different classification tasks with different prediction labels. Specifically, both vary in the \ac{BAC} thresholds that are classified as ``drunk driving''. Since there is no universal threshold across countries for when driving is forbidden, we base our \ac{ML} on the recommend legal limit by the \ac{WHO}, that is, a \ac{BAC} limit of \gdL{$0.05$}~\cite{whosafer2019}. Accordingly, the following definitions of labels were used for training and testing: (1)~early warning for predicting when the \ac{BAC} has already reached a \ac{MODERATE} \ac{BAC} level and (2)~above the limit for predicting when the \ac{BAC} is above the \ac{WHO} recommended legal limit~\cite{whosafer2019} of \gdL{$0.05$} (\ac{SEVERE}). We refer to the two classification tasks as \EARLY and \ABOVE. 

The predictive models for \ac{ML} were set to logistic regression with lasso regularization (i.e., L1 penalty). This choice was due to two reasons. First, logistic regression with lasso regularization results in parsimonious models, and, therefore, the risk of overfitting is comparatively low. Second, the model allows for straightforward interpretability, which allows us later to compare the model coefficients against prior knowledge on pathophysiological mechanisms.  

The \ac{ML} models were trained using log loss. We used the default implementation in Python 3.8 from the package scikit-learn (version 1.0.2)~\cite{scikit-learn}. We centered and standardized each feature wrt.\ to the training data using $z$-score normalization (centering with mean and dividing by \ac{SD}). Further, we set the class weights to be balanced and left the inverse regularization strength at the default value of 1.0. We tested the robustness of our hyperparameter choice in Appendix~\ref{app:robustness-checks}. Overall, our system remained robust to varying choices of lambda and regularization methods (i.e., L1 or L2 penalty).

\subsection{Model evaluation}

We evaluate the performance of our \ac{ML} system primarily based on the \ac{AUROC}~\cite{bestpracticesml2020}. The \ac{AUROC} has several benefits: it is widely used for classification tasks, considers the complete spectrum of decision thresholds, and accounts for class imbalances~\cite{weiss2013}. In addition, we report the \ac{AUPRC}. The \ac{AUPRC} is useful for settings where the positive class is particularly important (here: correct detection of drunk driving). Results for the \ac{AUPRC} are in Appendix~\ref{app:additional-results}. We also report further measures, namely balanced accuracy and F1 score (weighted by class). Here, we used a default $0.5$ probability for the decision threshold.

To evaluate our \ac{ML} system, we make use of \ac{LOSO} cross-validation~\cite{losocv}. Accordingly, a model is trained using the data from $n-1$ participants (i.e., all subsets except one) and then tested on the remaining $n$-th participant. This procedure is repeated for all $n$ participants. \ac{LOSO} cross-validation implies one important benefit: In contrast to standard cross-validation (i.e, $k$-fold within-participant cross-validation), \ac{LOSO} cross-validation evaluates the generalization capabilities of the \ac{ML} model to unseen participants~\cite{losocv}.

Results are reported as the out-of-sample prediction performance averaged across participants (i.e., macro-average). Further, the standard deviation is reported. This allows us to compare the variability in the performance across participants. Reassuringly, we remind that all hyperparameters are fixed and thus the same across models and participants to ensure generalizability.

\subsection{\ac{ML} baseline based on driving behavior only}
We further introduce a baseline based on \ac{CAN} data (i.e., all vehicle signals but without eye tracking) to evaluate the performance of our \ac{ML} system. In this evaluation, we apply the same feature processing pipeline as for the camera data. This pipeline also performed well in the past in detecting other driving states using~\ac{CAN} data, for example, detecting distraction~\cite{mcdonald2020}, emotions~\cite{liu2022}, low blood glucose levels~\cite{Lehmann2022}, and even intoxication~\cite{lee2010} of drivers. The simulator in our study captured~\ac{CAN} data regarding driver and vehicle behavior with a frequency of 30~Hz. Signals directly reflecting the driver behavior are the steering wheel angle as well as gas and brake pedal positions. For each of these signals, we further calculated the first and second derivative (i.e., velocity and acceleration). The vehicle behavior signals are the latitudinal as well as the longitudinal velocities and accelerations. Finally, we included the lateral position of the vehicle within the lane since related statistical analyses have described it as the most consistent factor in behavior change due to an alcohol intoxication~\cite{jongen2016}. We applied the same feature generation methods to the signals as for the camera data, that is, the same statistical aggregations and sliding window parameters of 60 seconds every 1 second.

\subsection{Post~hoc interpretability}

To interpret how the \ac{ML} systems arrive at predictions, we proceeded as follows. In line with existing research~\cite{npjlogisticregression2020a, npjlogisticregression2020b}, we assess the coefficients in our trained models to understand their underlying patterns. Features with a positive coefficient lead to a positive classification (i.e., a larger propensity to classify as drunk), whereas features with a negative coefficient lead to a negative classification (i.e., a lower propensity to classify as drunk). Moreover, as we normalize all features, the absolute size of each feature describes its importance on the output of the model. Hence, the coefficients explain the contribution of each feature to the model output.

%% file: body/04_results.tex
\section{Results}

In this section, we present the results of the evaluation of our machine learning system to detect drunk driving. First, we report the performance metrics of our machine learning system for the two classification tasks \EARLY and \ABOVE. Then, we compare our camera-based approach with a \ac{CAN} baseline, before examining the interpretability of our \ac{ML} system. Finally, we provide further insights into applicability by analyzing the decision time of our system.

\subsection{Performance evaluation}

\begin{figure*}[th]
    \centering
    \includegraphics[width=0.8\textwidth]{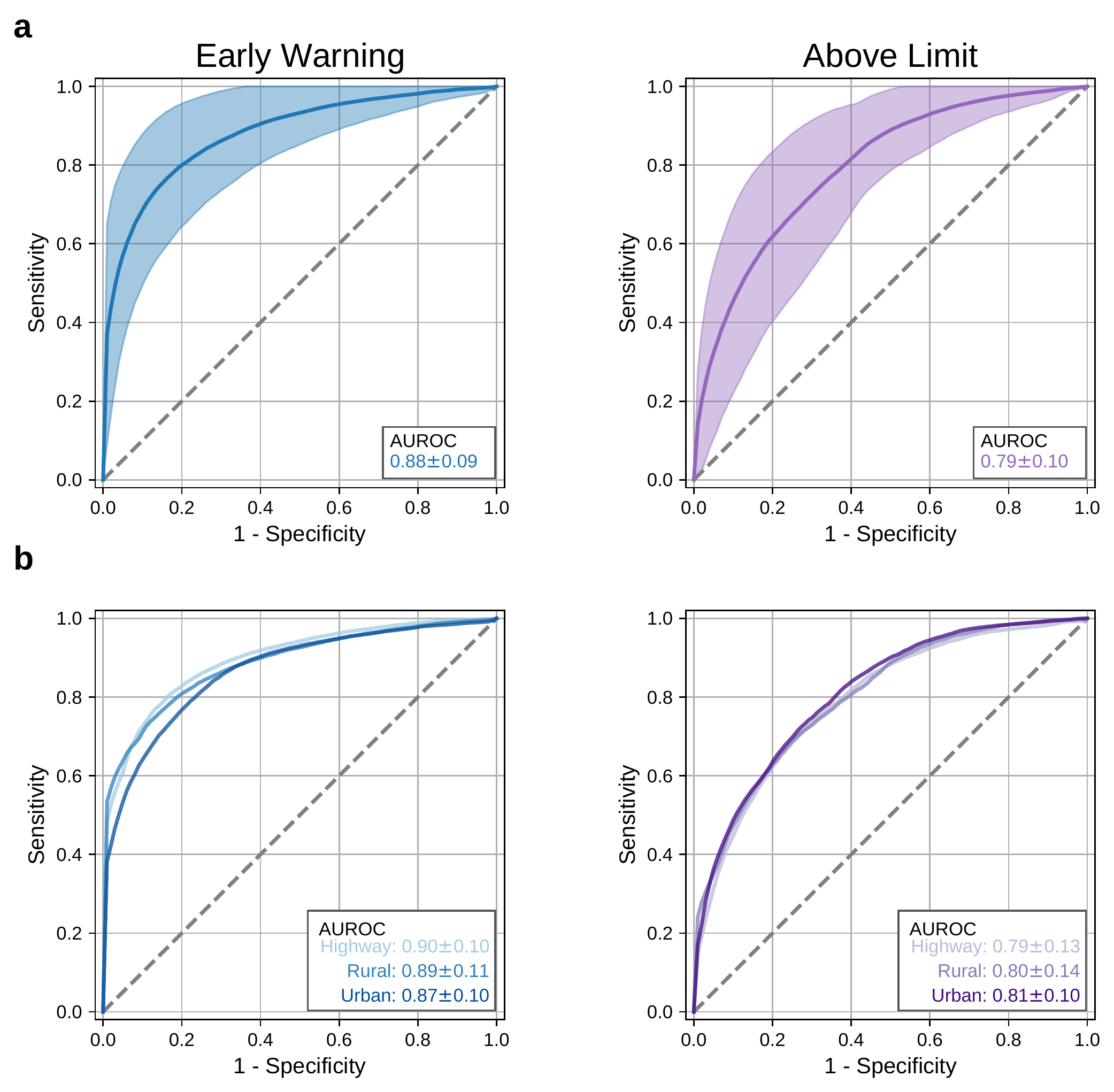}
    \Description{The figure shows the AUROC to evaluate the performance of the EARLY WARNING (left) and ABOVE LIMIT (right) machine learning systems. The two graphs in a) (top) show the overall performance. For the EARLY WARNING model, the overall performance is 0.88 ± 0.09. For the ABOVE LIMIT model, the overall performance is 0.79 ± 0.10. The two graphs in b) (bottom) show the performance per scenario. For the EARLY WARNING model, the performance for highway is 0.90 ± 0.10, for rural is 0.89 ± 0.11, for urban is 0.87 ± 0.10. For the ABOVE LIMIT model, the performance for highway is 0.79 ± 0.13, for rural is 0.80 ± 0.14, for urban is 0.81 ± 0.10.}
	\caption{\textbf{Performance of drunk driving detection.} The machine learning system for detecting drunk driving is evaluated based on the \acf{AUROC}. \textbf{(a)}~Performance across participants for different BAC thresholds. \textbf{(b)}~Performance by driving scenario (i.e., highway, rural, and urban). The dashed, gray line shows an AUROC of 0.50 as a na{\"i}ve baseline (i.e., a random guess). \acrexpl{BAC}.}~\label{fig:main_results}
\end{figure*}

The performance of the \ac{ML} system for detecting drunk driving is shown in \fref{fig:main_results}. The overall \acp{AUROC} for the two classification tasks with different \ac{BAC} thresholds are \meanstd{$0.88$}{$0.09$} (\EARLY) and \meanstd{$0.79$}{$0.10$} (\ABOVE). The respective \acp{SD} show that the performance across drivers is fairly stable. The \ac{ML} system further achieves a similar performance across different driving scenarios (highway, rural, and urban). For example, the mean \ac{AUROC} in the \EARLY task varies only between $0.87$ (urban) and $0.90$ (highway). We observe here and in the following analyses that the prediction performance is better for \EARLY than for \ABOVE but only to a small extent.

\begin{figure*}
    \centering
    \includegraphics[width=\textwidth]{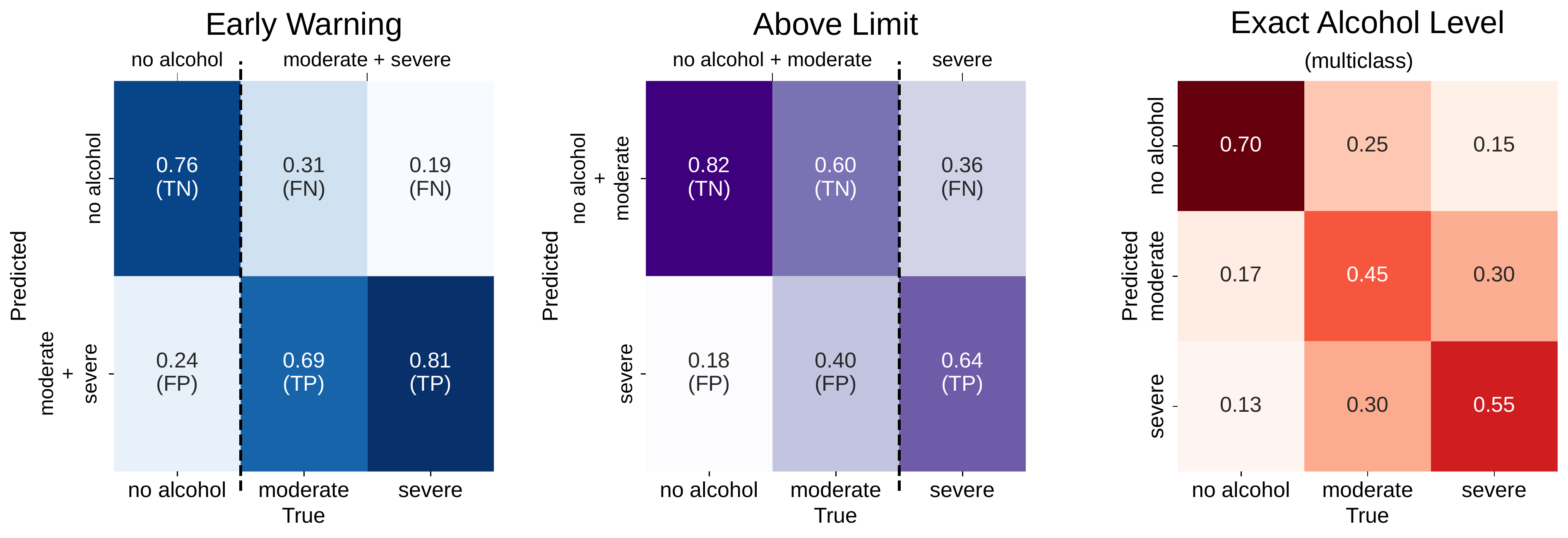}
\Description{The figure shows confusion matrices for EARLY WARNING (left), ABOVE LIMIT (middle), and for each exact alcohol level separate as a multiclass prediction (right) machine learning systems. The confusion matrices compare the relative frequency of the actual alcohol level (horizontal) vs. the predicted alcohol level (vertical). For comprehensiveness, we further report a granular breakdown for the actual alcohol level (horizontal) comparing no alcohol, moderate, and severe intoxication separately. For EARLY WARNING, the true negative rate is 0.84 (true: no alcohol, predicted: no alcohol), the false positive rate is 0.16 (true: no alcohol, predicted: any alcohol). The false negative rate is for true moderate and predicted no alcohol 0.27, for true severe and predicted no alcohol 0.15. The true positive rate is for true moderate and predicted any alcohol 0.73, for true severe and predicted any alcohol 0.85. For ABOVE LIMIT, the true negative rate for true no alcohol and predicted below the WHO recommended limit is 0.83, for true moderate and predicted below the WHO recommended limit is 0.56. The false positive rate for true no alcohol and predicted above the WHO recommended limit is 0.17, for true moderate and predicted above the WHO recommended limit 0.44. The false positive rate is 0.22 (true: below the WHO recommended limit, predicted: above the WHO recommended limit). The true positive rate is 0.78 (true: above the WHO recommended limit, predicted: above the WHO recommended limit). For the multiclass prediction, the system has a true positive rate of 0.70 for sober, 0.45 for moderate, and 0.55 for severe. Wrong classifications are rather low ranging between 0.13 and 0.25. The highest rate of wrong classification occurs when predicting moderate but the exact alcohol level is severe and when predicting severe but is actually moderate with 0.30 each.}
 \caption{\textbf{Confusion matrices for drunk driver detection.} The confusion matrices compare the relative frequency of the actual alcohol level (horizontal) vs. the predicted alcohol level (vertical). For comprehensiveness, we further report a granular breakdown for the actual alcohol level (horizontal) comparing no alcohol, moderate, and severe intoxication separately. The cells correspond to \acrexpl{FN}; \acrexpl{FP}; \acrexpl{TN}; \acrexpl{TP}. In addition, we provide a confusion matrix when our machine learning system is trained and evaluated on predicting each driving block/alcohol level separately.}~\label{fig:confusionmatrix}
\end{figure*}

Confusion matrices comparing the relative frequency of actual alcohol levels against predictions are shown in \fref{fig:confusionmatrix}. For \EARLY, both true positives and false negatives are comparatively infrequent, that is, the rate of false alarms and misses is low. A similar pattern is observed for \ABOVE prediction. Here, the rate of misses is again comparatively low (18\%). The rate of false alarms is also low for when drivers have no alcohol (14\%), while false positive tend to be more frequent when drivers have a moderate alcohol yet below the \ac{WHO} limit (41\%), implying that our \ac{ML} system is sensitive even to driving under little alcohol influence. This is further confirmed when our \ac{ML} model predicts each driving state separately. Here, we see that our model achieves a 70\% \ac{TP} rate for detecting the \ac{NOALCOHOL} driving state. Moreover, the two alcohol driving states have also a high \ac{TP} rate with 45\% (\ac{MODERATE}) and 55\% (\ac{SEVERE}), respectively, but with 30\% confusion between both intoxication states.

\subsection{Comparison of \ac{CAN} vs.\ camera approaches}

\begin{figure*}[!th]
  \centering
  \includegraphics[width=\textwidth]{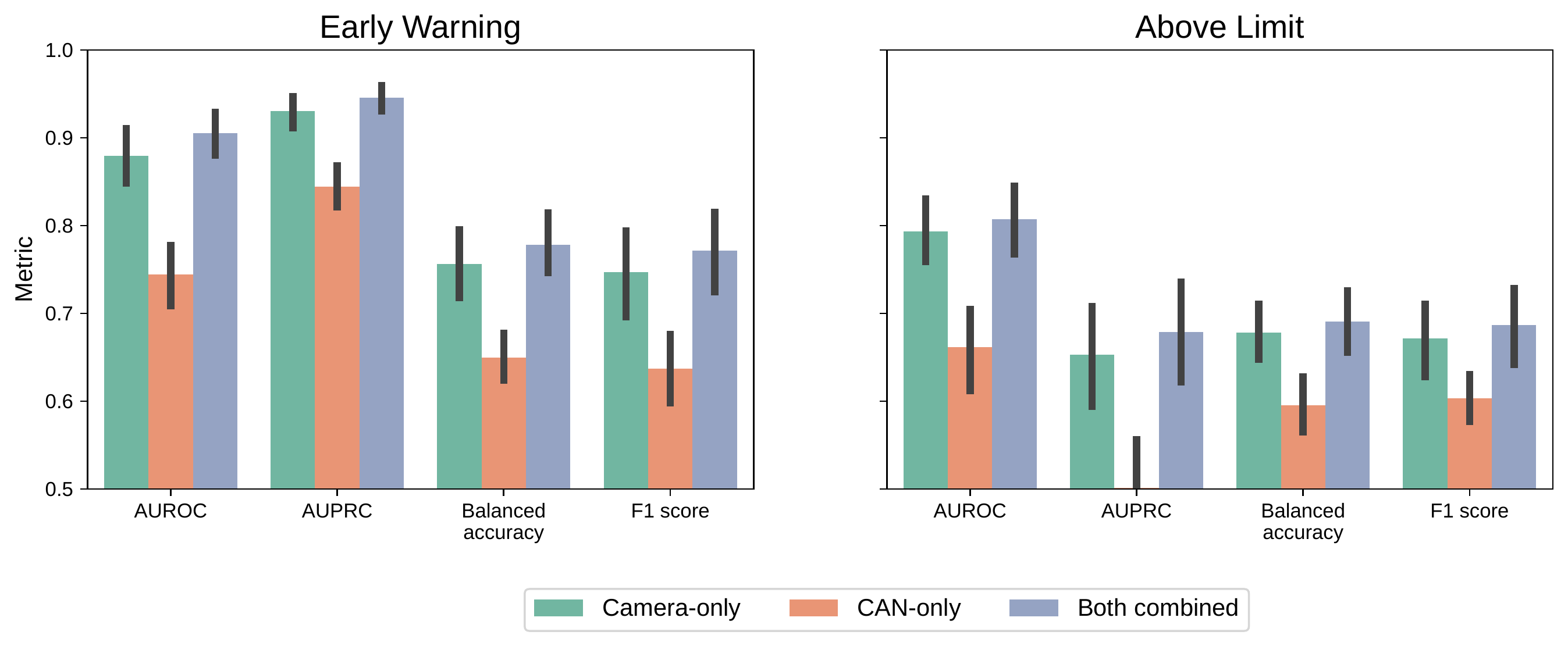}
  \Description{The figure shows prediction results for EARLY WARNING (left) and ABOVE LIMIT (right) machine learning systems using different data sources. CAN-only results are consistently lower than the camera-only or both data sources combined. For Early Warning, the following prediction results are achieved: AUROCs of 0.88 (camera-only), 0.74 (CAN-only), and 0.91 (both combined), AUPRCs of 0.93 (camera-only), 0.84 (CAN-only), and 0.95 (both combined), Balanced accuracies of 0.76 (camera-only), 0.65 (CAN-only), and 0.78 (both combined), and F1 scores of 0.75 (camera-only), 0.64 (CAN-only), and 0.77 (both combined). For Above Limit, the following prediction results are achieved: AUROCs of 0.79 (camera-only), 0.66 (CAN-only), and 0.81 (both combined), AUPRCs of 0.65 (camera-only), 0.50 (CAN-only), and 0.68 (both combined), Balanced accuracies of 0.68 (camera-only), 0.60 (CAN-only), and 0.69 (both combined), and F1 scores of 0.67 (camera-only), 0.60 (CAN-only), and 0.69 (both combined).}
  \caption{\textbf{Comparison of CAN vs.\ camera signals for detecting drunk driving.} We report the performance of our machine learning system while using different data sources: camera-only (our main system), CAN-only, and both combined. Overall, the results demonstrate the relevance of driver monitoring cameras. The following performance metrics are computed: \acf{AUROC}, \acf{AUPRC}, balanced accuracy, and F1 score (weighted by classes). Reported: \meanstd{mean}{\acl{SD}}. \acrexpl{CAN} (i.e., vehicle signals).}~\label{fig:comparison}%
\end{figure*}

Here, we compare our proposed camera-only approach against a \ac{CAN} baseline (i.e., vehicle signals only without camera data) and an approach combining both data sources (i.e., vehicle signals and camera data). The results are shown in \fref{fig:comparison}. In direct comparison to a baseline using CAN-only, our camera-only approach performs consistently better. For example, the CAN-only baseline achieves only an \ac{AUROC} of \meanstd{$0.74$}{$0.10$} (for \EARLY) and \meanstd{$0.66$}{$0.12$} (for \ABOVE), and thus has an \ac{AUROC} that is lower than that of the camera-only approach by around $0.10$. In line with this, the other performance metrics are also inferior for the CAN-only approach. For example, our camera-based \ac{ML} system records an \ac{AUPRC} of \meanstd{$0.93$}{$0.05$} (for \EARLY) and \meanstd{$0.65$}{$0.16$} (for \ABOVE). In contrast, the CAN-only baseline reaches an \ac{AUPRC} of only \meanstd{0.84}{0.07} (for \EARLY) and \meanstd{$0.50$}{$0.15$} (for \ABOVE). Furthermore, the camera-only \ac{ML} system achieves a balanced accuracy of \meanstd{$0.76$}{$0.10$} (for \EARLY) and \meanstd{$0.68$}{$0.10$} (for \ABOVE), whereas the CAN-only baseline results in a balanced accuracy of only \meanstd{$0.65$}{$0.08$} (for \EARLY) and \meanstd{$0.60$}{$0.09$} (for \ABOVE). Comparable results are achieved for the F1 score, where the camera-only system achieves \meanstd{$0.75$}{$0.14$} (for \EARLY) and \meanstd{$0.67$}{$0.12$} (for \ABOVE). Again, the CAN-only baseline is substantially outperformed. The latter registers an F1 score of \meanstd{$0.64$}{$0.11$} (for \EARLY) and \meanstd{$0.60$}{$0.08$} (for \ABOVE).

The camera-only approach achieves a performance similar to that of an approach combining both \ac{CAN} and camera data. As an example, the \ac{ML} system combining both data sources achieves an \ac{AUROC} of \meanstd{$0.91$}{$0.07$} (for \EARLY) and \meanstd{$0.81$}{$0.11$} (for \ABOVE). In comparison, the camera-only \ac{ML} system records \meanstd{$0.88$}{$0.09$} (for \EARLY) and \meanstd{$0.79$}{$0.10$} (for \ABOVE). In sum, using CAN instead of camera signals is inferior and combining both CAN and camera signals does not lead to a statistically significant improvement. As such, the results corroborate the relevance of driver monitoring cameras to detect drunk driving.

\subsection{Robustness checks}

\begin{figure*}
    \centering
    \includegraphics[width=0.85\textwidth]{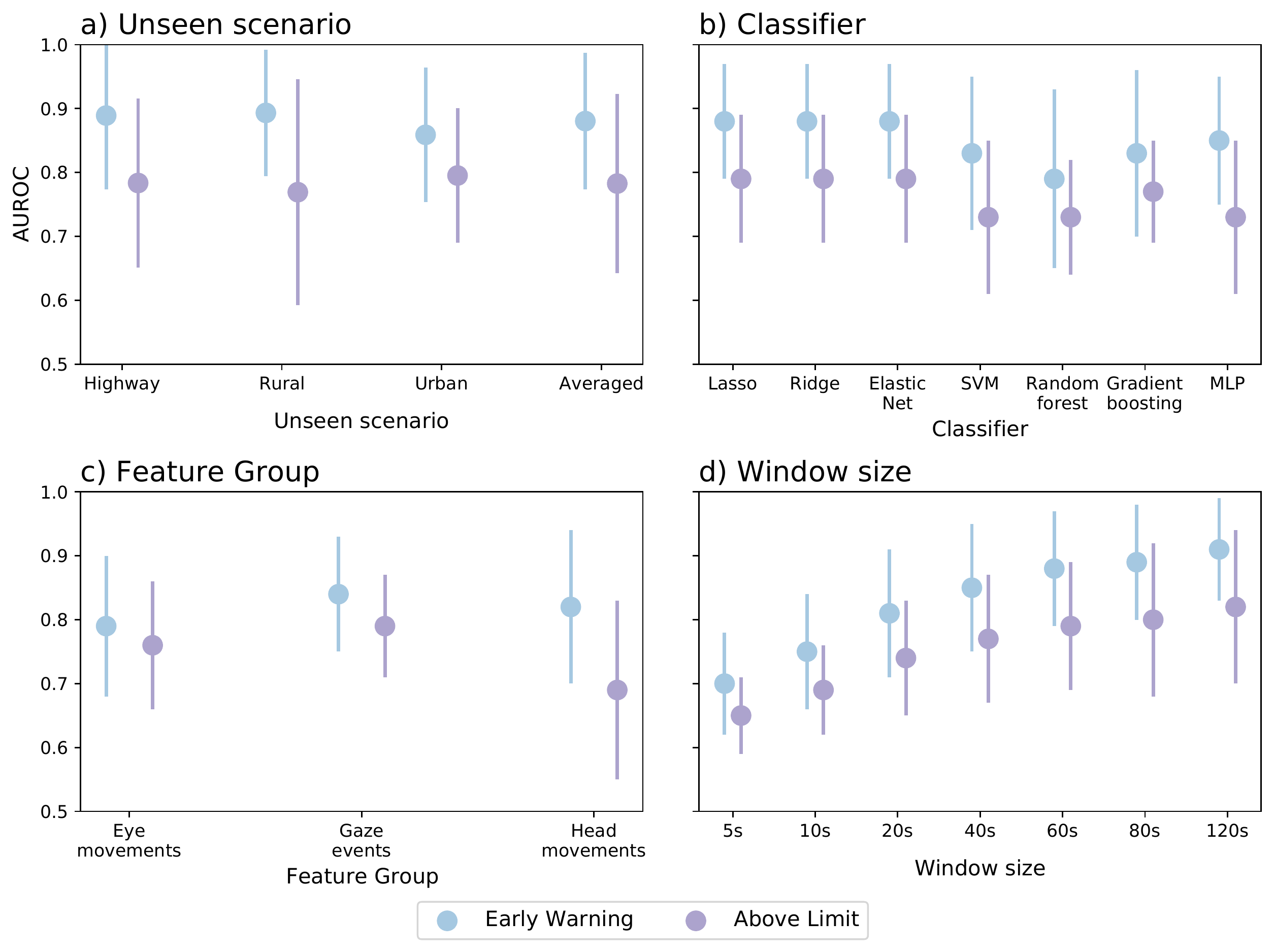}
    \Description{Four graphs showing different sensitivity analyses based on the AUROC performance for the drunk driver detection. Graph a) shows the sensitivity analysis when in training one scenario is left out and then evaluated on it. The results are similar to the validation of our system with leave-one-subject-out alone, except that the standard deviation slightly increases. The AUROC results are 0.89 (EARLY WARNING) and 0.78 (ABOVE LIMIT) for highway, 0.94 (EARLY WARNING) and 0.77 (ABOVE LIMIT) for rural, 0.92 (EARLY WARNING) and 0.80 (ABOVE LIMIT) for urban, and 0.82 (EARLY WARNING) and 0.78 (ABOVE LIMIT) for all scenarios averaged. Graph b) shows the sensitivity analysis for different classifiers. Logistic regressions with different regularizations (lasso, ridge, elastic net) perform best with AUROC mean results of 0.88 (EARLY WARNING) and 0.79 (ABOVE LIMIT). SVM, Gradient boosting, Random Forest, and MLP still perform well with AUROC mean results above 0.79 (EARLY WARNING) and 0.73 (ABOVE LIMIT), however, they have larger standard deviations. Graph b) shows the sensitivity analysis for different feature groups. Overall gaze events perform best with AUROC mean results of 0.85 (EARLY WARNING) and 0.79 (ABOVE LIMIT). It is followed by eye movements with AUROC mean results of 0.79 (EARLY WARNING) and 0.76 (ABOVE LIMIT) and head movements with AUROC mean results of 0.81 (EARLY WARNING) and 0.69 (ABOVE LIMIT). Notably, head movements feature have a larger standard deviation than both other feature groups. Graph c) shows the sensitivity analysis for different window sizes. The mean mean AUROC results are for window sizes of 5 seconds 0.70 (EARLY WARNING) and 0.66 (ABOVE LIMIT), of 10 seconds 0.75 (EARLY WARNING) and 0.69 (ABOVE LIMIT), of 20 seconds 0.82 (EARLY WARNING) and 0.72 (ABOVE LIMIT), of 40 seconds 0.84 (EARLY WARNING) and 0.76 (ABOVE LIMIT), of 60 seconds 0.88 (EARLY WARNING) and 0.79 (ABOVE LIMIT), of 80 seconds 0.89 (EARLY WARNING) and 0.80 (ABOVE LIMIT), of 120 seconds 0.90 (EARLY WARNING) and 0.80 (ABOVE LIMIT).}
	\caption{\textbf{Sensitivity analysis for drunk driver detection.} The machine learning system for detecting drunk driving is evaluated based on the \acf{AUROC} while varying different components of our \ac{ML} system: \textbf{(a)}~leaving one scenario in training out and testing on it (i.e., leave-one-subject and leave-one-scenario out cross validation), \textbf{(b)}~the classifier, \textbf{(c)}~the features, and \textbf{(d)}~the size of the sliding window. Across all configurations, the machine learning system has a robust, high prediction performance. \acrexpl{MLP}; \acrexpl{SVM}.}~\label{fig:sensitivity}
\end{figure*}

Additional sensitivity analyses were performed to evaluate the robustness of the \ac{ML} system (see \fref{fig:sensitivity} and Appendix~\ref{app:robustness-checks}). First, we evaluated how our \ac{ML} system performs on unseen driving scenarios. Therefore, we introduced a leave-one-driving-scenario-out cross-validation on top of our \ac{LOSO} cross-validation. More specifically, we performed three evaluations for each participant: each driving scenario was omitted once in training and evaluation was based on the left-out driving scenario. Our \ac{ML} system achieves similar results as with the \ac{LOSO} cross-validation alone, showing that our system works independent of driver and scenario. Second, we varied the length of the sliding window for feature engineering. Overall, the results of our \ac{ML} system remain robust. We observe a tendency that longer sliding windows are associated with a larger \ac{AUROC}. However, informed by literature on detecting other critical driving events~\cite{drowsiness_similar_pipeline2019}, we set the window length to 60 seconds in the above analyses, as it allows for timely feedback. Second, we compared the predictive power of the different features from eye movements, gaze events, and head movements. Here, we observe a larger \ac{AUROC} for gaze events (i.e., fixations and saccades), followed by eye movements (i.e., velocity and acceleration) and head movements (i.e., velocity and acceleration). Across all, the mean \ac{AUROC} remains consistently above $0.70$. Third, we repeated the analysis with alternative \ac{ML} models. We used both linear models (logistic regression with lasso, ridge, and elastic net regularization) and non-linear models (support vector machine, random forest, gradient boosting model, and multi-layer perceptron). Overall, the logistic regression with lasso regularization performs best and thus justifies our model choice. Nevertheless, all models achieved a mean \ac{AUROC} of 0.73 or better, which corroborates the robustness of our prediction.

\subsection{Interpretability}

\begin{figure*}[th]
    \centering
    \includegraphics[width=0.85\textwidth]{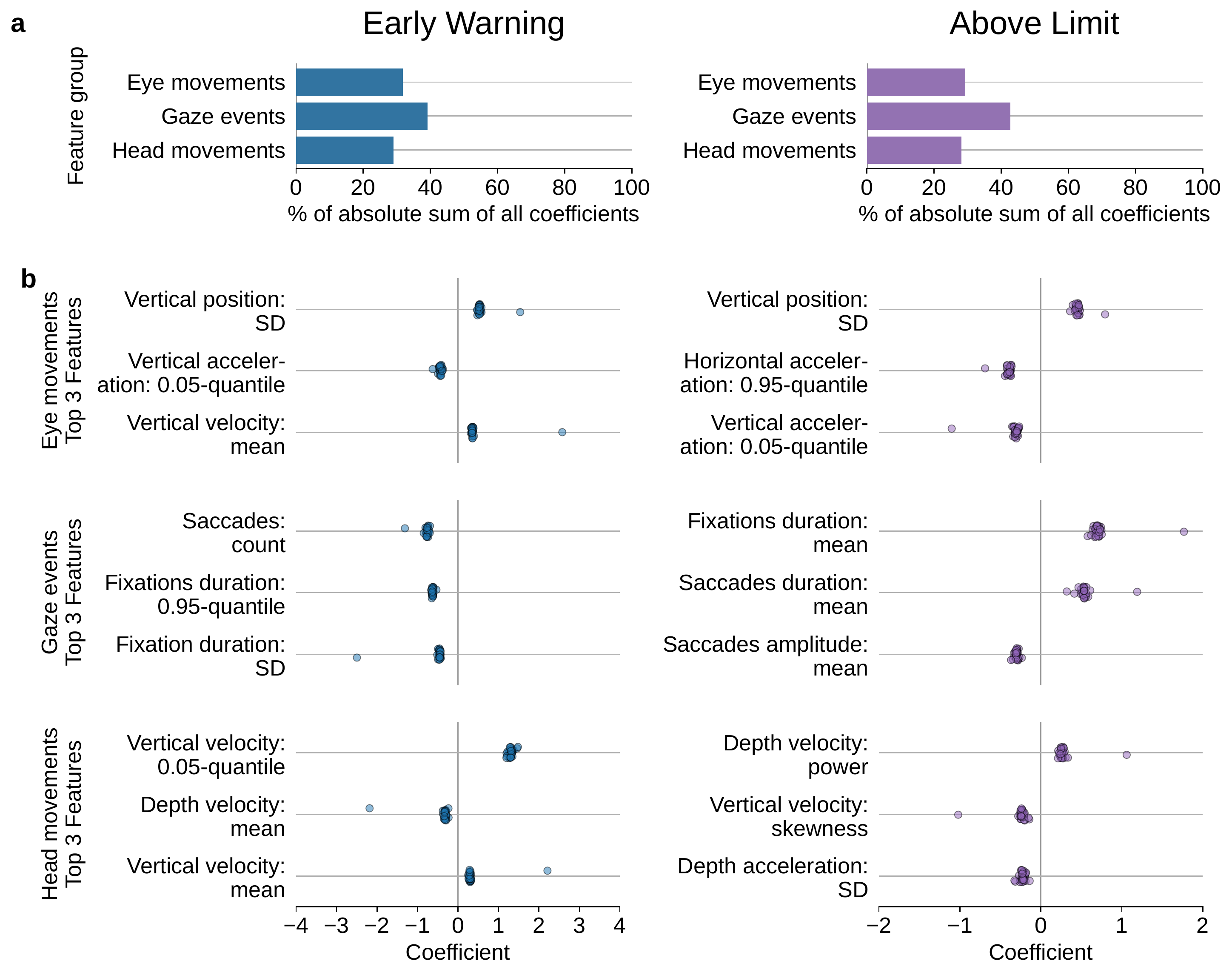}
	\Description{The figure shows graphs for interpretability of the EARLY WARNING (left) and ABOVE LIMIT (right) machine learning systems. The two graphs in a) (top) show the overall importance of different feature groups (eye movements, gaze events, and head movements). In both models, the gaze event features have in total the highest value, followed by eye movements and head movements. The two graphs in b) (bottom) show the top-3 features in each feature group. For each feature, the coefficients from each trained model of the leave-one-subject-out cross-validation are shown. In the following, we will list the top-3 features for each feature group for each classification with its mean coefficients across all cross-validation models. Notably, the coefficients are rather stable across all cross-validation runs. For EARLY WARNING, the top-3 features for eye movements are vertical position: SD with a mean coefficient of 0.8, vertical acceleration: 0.05-quantile with a mean coefficient of -0.5, and vertical velocity: mean with a mean coefficient of 0.5. For ABOVE LIMIT, the top-3 features for eye movements are vertical position: SD with a mean coefficient of 0.8, horizontal acceleration: 0.95-quantile with a mean coefficient of 0.3, and vertical acceleration: 0.05-quantile with a mean coefficient of 0.3. For EARLY WARNING, the top-3 features for gaze events are saccades: count with a mean coefficient of -0.8, fixations duration: 0.95-quantile with a mean coefficient of -0.7, and fixation duration: SD with a mean coefficient of -0.5. For ABOVE LIMIT, the top-3 features for gaze events are fixations duration: mean with a mean coefficient of 0.7, saccades duration: mean with a mean coefficient of 0.6, and saccades amplitude: mean with a mean coefficient of -0.3. For EARLY WARNING, the top-3 features for head movements are vertical velocity: 0.05-quantile with a mean coefficient of 1, depth velocity: mean with a mean coefficient of -0.2, and vertical velocity: mean with a mean coefficient of 0.15. For ABOVE LIMIT, the top-3 features for head movements are depth velocity: power with a mean coefficient of 0.2, vertical velocity: skewness with a mean coefficient of -0.15, and depth acceleration: SD with a mean coefficient of -0.15.}
 \caption{\textbf{Interpretability for machine learning.} The coefficients are analyzed in order to examine how the logistic regression arrives at predictions. \textbf{(a)}~The importance of different feature groups (eye movements, gaze events, and head movements) is compared. Formally, we sum over the absolute values of the coefficients in each group and then normalize them to one. This yields the \% of the sum of the coefficients (absolute value) per feature group and thus quantifies the relative importance. \textbf{(b)}~The regression coefficients of the top-3 features in each feature group are reported. Separate dots ($n=30$) are shown for the coefficients from the different splits during cross-validation. \acrexpl{SD}.}~\label{fig:coefficients}
\end{figure*}

To explain the decision logic in the \ac{ML} model, we interpret the coefficients and thereby assess how features are associated with the predictions. First, we focus on the absolute magnitude of the coefficients  (\fref{fig:coefficients}a), which allows us to identify features that are important for the \ac{ML} model. Here, we group features by eye movements (velocity, acceleration), gaze events (fixations, saccades), and head movements (velocity, acceleration). Overall, features from gaze events receive coefficients with larger absolute value than the other feature groups and are thus highly important for the \ac{ML}. Moreover, for detecting an alcohol intoxication, eye movements are more important than head movements. We also evaluated each feature group separately for their predictive performance (see Appendix~\ref{app:additional-results}). 

The model coefficients are shown in \fref{fig:coefficients}b. For example, the \ac{ML} model for \ABOVE is ceteris paribus more likely to predict drunk driving when drivers have a longer mean duration of fixations and saccades and a shorter mean amplitude (i.e., distance traveled) of saccades.

\subsection{Decision time}
Decision time plays a critical role in achieving timely as well as reliable decisions for our system before interventions are delivered. To examine the decision time, we computed the balanced accuracy for each second of a trip. To do this, we applied a rolling cumulative moving average to the predicted probability of each window in each separate trip across all drivers. The predictions reach a high plateau early on. After 90 seconds of driving, this approach already achieves a balanced accuracy with 95\% \ac{CI} of $0.77$ [$0.72$, $0.82$] (\EARLY) and $0.69$ [$0.64$, $0.75$] (\ABOVE). The decision frequency of our system is every second, and, hence, a filter is needed to prevent a potentially volatile decision behavior and assure stable and reliable intervention delivery. Therefore, we evaluated a non-overlapping majority vote for the predicted windows (such as in~\cite{recofit2014}). The prediction performance of our \ac{ML} system reaches a peak at the aggregation of 150 windows with an \ac{AUROC} of \meanstd{$0.91$}{$0.08$} (\EARLY) and \meanstd{$0.85$}{$0.11$} (\ABOVE). Both analyses show that our \ac{ML} system provides timely and reliable decision after only a few minutes.

\begin{figure*}[th]
    \centering
    \includegraphics[width=0.85\textwidth]{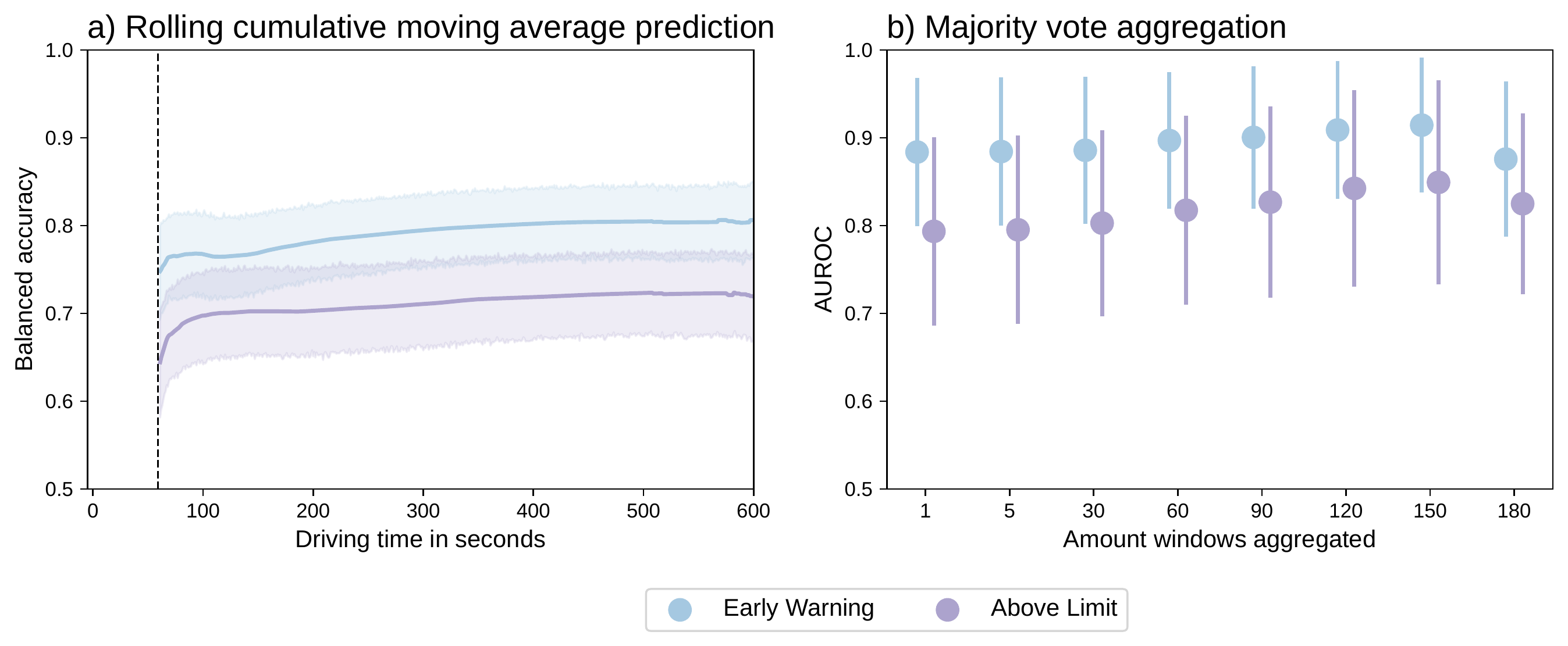}
	\Description{The figure shows two graphs for approaches to understand potential decision-making approaches. Left is a figure that shows the application of a rolling cumulative moving average on the predicted windows of both classification tasks (EARLY WARNING and ABOVE LIMIT). The x-axis shows the driving time in seconds, the  y-axis the balanced accuracy. After 60 seconds, the first prediction starts. For the two classification tasks, the balanced accuracy is reported as a line including 95\% confidence intervals. The balanced accuracies start at 0.73 (EARLY WARNING) and 0.65 (ABOVE LIMIT). Already after 90 seconds, the aggregation leads to balanced accuracies of 0.77 (EARLY WARNING) and 0.69 (ABOVE LIMIT). Until the end of the trips (600s), the balanced accuracies rise to 0.80 (EARLY WARNING) and 0.71 (ABOVE LIMIT). The right figure shows the results of a non-overlapping majority vote aggregations for different window aggreations. On the x-axis, the different window aggregation sizes with no, 5, 30, 60, 90, 120, 150, and 180 windows aggregated. On the y-axis, the AUROC is visible. The AUROC are decreasing until an aggregation of 150 windows. No aggregation has an AUROC of 0.88 (EARLY WARNING) and 0.79 (ABOVE LIMIT). At 150 windows aggregated, the system achieves an AUROC of is at 0.91 (EARLY WARNING) and 0.85 (ABOVE LIMIT).}
 \caption{\textbf{Insights into the decision-making time for the machine learning system for drunk driving detection.} Drunk driving detection needs fast and stable decision-making, thus the decision time and frequency is evaluated. \textbf{(a)}~Decision time and the performance when applying a cumulative moving average to the predicted probability of each window along the driving time. \textbf{(b)}~Decision time and the performance when applying a non-overlapping majority vote aggregation of the predicted windows. \acrexpl{AUROC}.}~\label{fig:decision}
\end{figure*}

%% file: body/05_discussion.tex
\section{Discussion}

\subsection{Contributions}

Alcohol consumption is responsible for a major share of the global disease burden and overall mortality~\cite{whoroadsafety2018, whoalcohol2019}. There is thus a need for scalable and cost-effective \ac{HCI} technology towards behavioral change with the aim of reducing alcohol-related harms. Here, we developed and evaluated a novel machine learning system for drunk driver detection based on driver monitoring cameras. To the best of our knowledge, our system is the first to detect drunk driving from camera-based sensor technology. 

Our system achieves a high detection performance with an \ac{AUROC} of $0.88$ (for detecting driving with a \ac{BAC} $>$ \gdL{$0.00$} and $\leq$ \gdL{$0.03$}) and $0.79$ (for detecting driving above the \ac{WHO} recommended legal limit of \gdL{$0.05$}) while testing with a leave-one-subject-out cross-validation. Moreover, we achieve similar performances when applying a leave-one-subject-out and leave-one-driving-scenario-out cross-validation with an \ac{AUROC} of $0.82$ (\ac{BAC} $>$ \gdL{$0.00$} and $\leq$ \gdL{$0.03$}) and $0.78$ (above the \ac{WHO} recommended legal limit of \gdL{$0.05$}). Hence, the results demonstrate that our system can even provide early warnings when there is only a moderate alcohol intoxication.

Our \ac{ML} system relies upon driver monitoring cameras as input. Driver monitoring cameras are nowadays common in modern vehicles as part of driver assistance systems. Moreover, driver monitoring cameras are to become mandatory for all new vehicles due to safety regulations. Examples of such regulations are the \ac{Euro NCAP} and the \EU \ac{GSR}, which make driver monitoring cameras mandatory from 2024 onwards~\cite{encaproadmap2017, gsr2019}. Notwithstanding, other technologies such as breath-based sensors~\cite{dadds2019} allow for the detection of drunk driving. However, the current state of such technologies is expensive and requires regular maintenance~\cite{radun2014, swov2005}. In contrast to that, the growing availability of driver monitoring cameras makes them a scalable, low-cost, and easily accessible technology. 

\subsection{Comparison with previous work}
Previous works on detecting drunk driving are based on driving behavior (e.g., steering, pedal usage, vehicle speed)~\cite{lee2010,drunk_driving_related_1, drunk_driving_related_2, drunk_driving_related_3, drunk_driving_related_4, drunk_driving_related_5, drunk_driving_related_6, drunk_driving_related_7, drunk_driving_related_8, drunk_driving_related_9}. However, to the best of our knowledge, no work has so far developed or evaluated a system based on driver cameras. This is our novelty.

Here, we propose to shift from driving behavior to driver behavior and, specifically, to leverage eye gaze and head movements. Our choice has important benefits. (1)~Visual and perceptual impairments due to alcohol already occur at a \ac{BAC} of \gdL{0.005}. In comparison, changes in vehicle control occur only for a much larger \acp{BAC} \cite{jongen2016, moskowitz2000}. This thus makes predictions from eye gaze particularly effective for early warnings of drunk driving. (2)~Driver monitoring cameras have recently been applied to a related task, namely detecting driver drowsiness \cite{drowsiness2015, drowsinesssystems2020}. Here, predictive gains have been achieved by moving from driving behavior to driver behavior, analogous to our work. (3)~Driving behavior (e.g., steering, pedal usage, vehicle speed) is subject to large variability across driving scenarios (highway, rural, and urban) and across drivers (e.g., slow vs. fast drivers, defensive vs. aggressive drivers) and thus overlays the raw signals with noise~\cite{lee2010}.

Our work shows that the predictive power of eye gaze and head movements is robust and can reliably generalizes across participants and to unseen driving scenarios. We found that the camera-based approach clearly outperforms a CAN-only baseline as used by related work (e.g.,~\cite{lee2010}). For example, the CAN-only baseline achieves only an \ac{AUROC} of \meanstd{$0.74$}{$0.10$} (for \EARLY) and \meanstd{$0.66$}{$0.12$} (for \ABOVE), and thus has an \ac{AUROC} that is lower than that of the camera-only approach by around $0.10$. Combining both camera and CAN, we achieve similar performances. This underlines that the camera information are the major factor in detecting drunk driving. Moreover, this hypothesis is confirmed when we compare our prediction performance with different previous studies. For example, in one study, the \US \ac{NHTSA} examined the extent to which driving below and above a \ac{BAC} of \gdL{0.08} (the legal limit in the \US) can be detected~\cite{lee2010}. Despite the higher alcohol threshold, we can put into context the results with our study because the study design, driving task, and analysis are comparable to ours. For the same driving scenarios, the work based on driving behavior (e.g., steering, pedal usage) achieves an \ac{AUROC} of \meanstd{0.77}{0.08}. Thereby, data from all drivers in sober and intoxicated states is used. For comparison, our system achieves a better prediction performance for a lower \ac{BAC} threshold of \gdL{0.05} (the \ac{WHO} recommended limit) with an \ac{AUROC} of \meanstd{0.79}{0.10} validated out-of-subject. In essence, this demonstrates the effectiveness of using driver monitoring cameras for detecting drunk driving.

\subsection{Interpretations of machine learning model in relation to pathophysiology}

The post~hoc interpretation of our models is in line with known pathophysiological effects that describe impairments induced by alcohol. Examples of such pathophysiological effects are divided attention, lower reaction times, changes in vigilance, tracking, perception, and psychomotor functions~\cite{moskowitz2000}. As we see in our analysis, these pathophysiological effects are especially pronounced for {gaze events}. Our system learns to recognize that, with increased intoxication, the mean duration of fixations increases and fixation frequencies decrease. This is consistent with pathophysiological effects according to which people under the influence of alcohol take longer time to process visual information~\cite{moskowitz2000, fixation2020}. Moreover, we observe changes in saccadic eye movements (e.g., fewer and longer saccades). This can be explained by prior literature on visual scanning \cite{saccades2010, saccade2013, shiferaw2019}. Specifically, drunk drivers tend to follow scattered and irregular patterns of visual sampling, which is reflected by changes in saccade velocities and amplitudes~\cite{saccades2010, saccade2013, shiferaw2019}. In sum, the observed changes in fixations and saccades are consistent with existing research and well-described phenomena such as tunnel vision~\cite{moskowitz2000, tunnelvision2001}. 

\subsection{Time until a decision}
Most alcohol-related road crashes take place on free, straight roads (i.e., out of city, higher speed, and no curves)~\cite{destatis2020}; therefore, our system needs to make a reliable decision before drivers reaches such roads. In our evaluation (see~\fref{fig:decision}), we find that our system achieves a confident decision already after around 90 seconds (balanced accuracy with 95\% \ac{CI} of $0.77$ [$0.72$, $0.82$] (\EARLY) and $0.69$ [$0.64$, $0.75$] (\ABOVE)). Moreover, a majority vote aggregation would further increase the reliability. The prediction performance of our \ac{ML} system reaches a peak at the aggregation of 150 windows with an \ac{AUROC} of \meanstd{$0.91$}{$0.08$} (\EARLY) and \meanstd{$0.85$}{$0.11$} (\ABOVE). Detecting the alcohol level before driving would be the safest and most reliable way in preventing drunk driving. However, current technology such as alcohol ignition locks are highly expensive and require regular maintenance so that a widespread application of this preventive technology is highly unlikely ~\cite{radun2014, swov2005}. Given the number of alcohol-related driving incidents, waiting for the highly effective and affordable ``silver bullet'' to prevent drunk driving seems rather unethical. Instead of insisting on a perfect solution and waiting for years until manufacturers and/or regulators decide on a sensor technology that has yet to be developed and evaluated, smaller steps can be taken as of today. Our approach relies on existing technologies of modern vehicles, which allows a swift introduction of measures against the problem of drunk driving.

\subsection{Limitations}
The strength of this study is the rigorous development and evaluation of a \ac{ML} system for the detection of drunk driving using driver monitoring cameras. For this, we performed an interventional study based on a standardized procedure for alcohol administration~\cite{bestpractices2019, bestpractices2008, bestpractices2013} and, further, followed best practice in \ac{ML}~\cite{hastie2009}. Nevertheless, this study has limitations.

\subsubsection*{Environment and driving-related factors}
First, as with all simulator studies, there may be risk of latent effects, such as learning or drowsiness. However, we follow best practice for simulator studies to prevent such latent effects to the extent possible~\cite{bestpractices2019, bestpractices2008, bestpractices2013}. Specifically, we mitigated learning effects by giving participants sufficient time for practicing before the start of the driving tasks and reduced drowsiness by introducing comparatively long breaks between each driving block. Second, our study examined driving in a simulator rather than in a real vehicle. In real-world driving, drivers may adapt their visual activity to the environment (e.g., at night or in rain) or perform secondary tasks while driving (e.g., talking to passengers or making phone calls). Real-world driving studies are recommended to address these issues, and we discuss how these issues should be addressed in future work (see~Section~\ref{sec:outlook}). However, previous work suggests that simulators reliably reproduce changes in driver behavior under the influence of alcohol~\cite{jongen2016, helland2013}. By covering a wide range of scenarios and traffic situations, we demonstrated the generalizability of our approach to different driving situations. This is underlined by the stable results of the leave-one-driving-scenario-out cross-validation. Finally, we also want to highlight that our results rely on camera technology that produces reliable results even in demanding situations. Our system is based on an infrared camera (as also used in industry~\cite{bosch_dms2022}). These cameras work in various light conditions and still provide reliable results even in situations with strong light changes, night driving or even when drivers wear (sun)glasses.

\begin{figure*}
\centering
    \includegraphics[width=\textwidth]{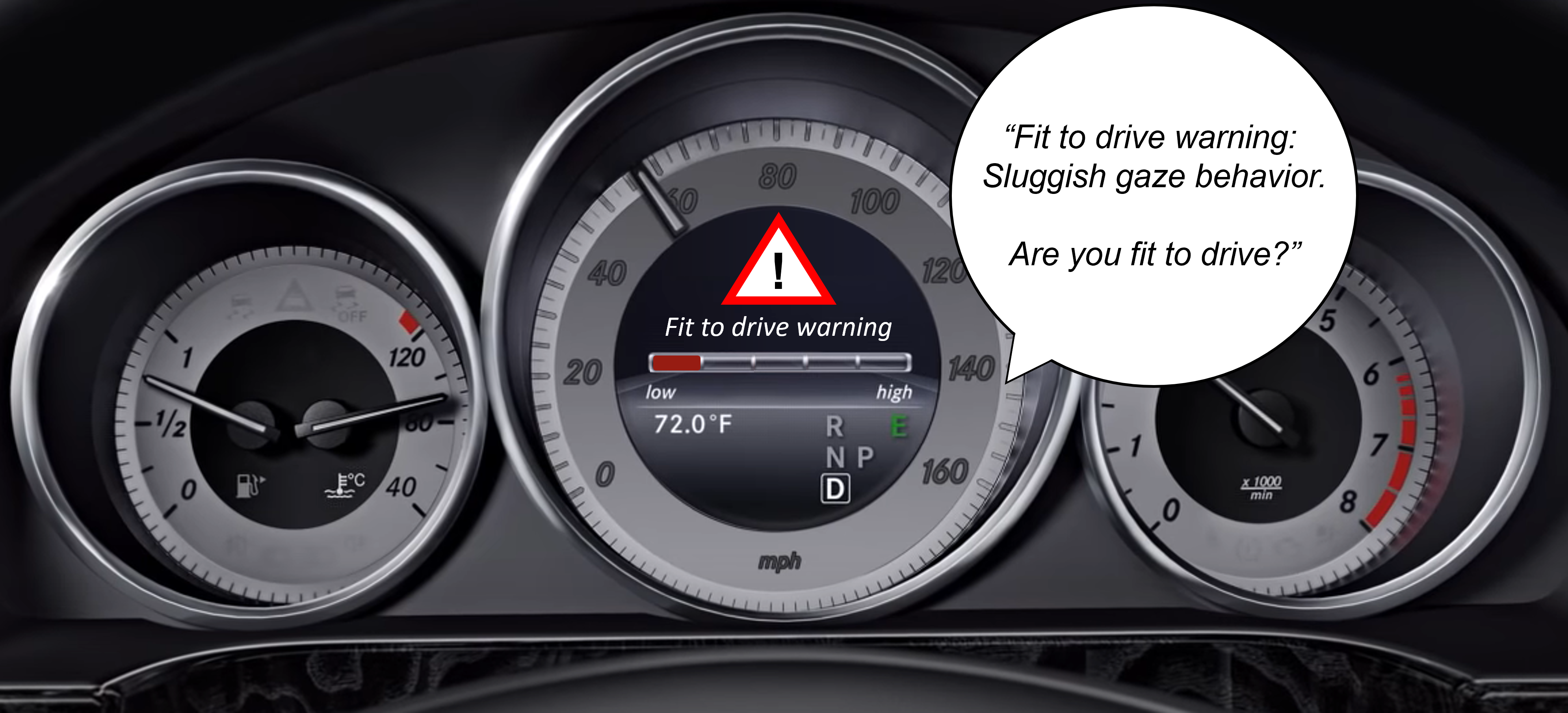} 
\Description{This figure shows a car instrument cluster and an example warning based on our system. On the car instrument cluster is an example of a warning shown. The warning consists of a warning symbol, a text message (``Fit to drive warning''), and a vertical bar ranging from low to high with a point close to low. Moreover, a speech bubble is visible that indicates a warning message stating: ``Fit to drive warning: Abnormal sluggish gaze behavior. Are you fit to drive?''.}
\caption{\textbf{Example for a fit to drive intervention system.} The system warns the driver with a visual- and audio-based intervention including a reason for the warning.}
\label{fig:intervention}
\end{figure*}

\subsubsection*{Individual driver factors}
In our study, we were able to show that our system is able to have a high detection accuracy for a demanding leave-one-subject-out cross-validation across different age groups and genders. Therefore, our system is capable of handling individual tolerance, at least for our study population, which consisted of healthy individuals from Switzerland with regular alcohol consumption. We see two potential barriers why our system performance could potentially deteriorate when applying it to other populations. First, due to individual alcohol tolerances, the effect of alcohol levels can potentially vary from person to person~\cite{tolerance2013}. For example, different ethnicities have a different sensitivity to alcohol~\cite{ethnic1972}. However, the \ac{WHO} recommendation of a \ac{BAC} of \gdL{0.05} as the legal limit for driving under the influence is chosen to reflect in which the majority of people show signs of impaired driving~\cite{alcohol_limit_005}. Even experienced drinkers show impairments in their driving behavior as early as a \ac{BAC} of \gdL{0.02}~\cite{moskowitz2000}. Therefore, our current system should already cover the problem of individual tolerances even for non-included healthy populations (e.g., a different ethnicity). Second, our system may make incorrect classifications for populations with different visual scanning behaviors. Most likely, the underlying gaze detection algorithm of our driver monitoring system will fail for individuals with health problems that affect their eye and pupil movements, such as strabismus or nystagmus. Very young or very old drivers might also challenge our system. Young drivers have less situation-aware visual activity due to their limited driving experience~\cite{young_drivers2005}, whereas older people have slower visual processing times than other age groups~\cite{senior_gaze2013}. Additional experiments are needed for these two populations. However, specifically in the case of elderly people, our system could create value by identifying fundamental and safety-critical changes in behavior. In the following, we will discuss possible interventions to support drivers.

\subsection{Implications}

The accurate detection of moderate alcohol levels by our system is an important prerequisite for providing effective digital interventions to prevent alcohol-related harm~\cite{shani2015, shani2017, selfmonitoringalcohol2015}. This is especially relevant as people regularly fail to correctly self-assess their alcohol levels~\cite{nobuzz2016, npjalcohol2021}. Drivers consistently underestimate their alcohol intoxication and therefore overestimate their ability to drive. Here, our system could be used to trigger behavioral interventions. One example is drunk driving warnings which promote transparency similar to self-tracking and thus train people in \ac{BAC} discrimination~\cite{selfestimation2013} and drinking control strategies~\cite{intervention_drink_less}.

\fref{fig:intervention} illustrates a potential future warning intervention. We envision a comprehensive driver warning system that conducts a fit to drive assessment. Such a system could address drunk driving but also other impairments, for example, drowsiness or lack of attention. In addition, driving under the influence of cannabis, ecstasy, or other illicit drugs could be also included in the future as they are known to impair the driving behavior as well~\cite{drowsinesssystems2020, cannabis2004, mdma2006}. Improved transparency has been shown to be a key driver of behavioral change in the context of self-tracking~\cite{selfestimation2013, SoberDiary2015, selfmonitoringalcohol2015}. Accordingly, the system could inform the driver of the observed behavior that led to the warning. In the example of \fref{fig:intervention}, the machine learning system identified longer fixation times and thus slower information processing of the driver as the reason for the warning.

Beyond warnings, there are also more restrictive interventions for escalation. For example, modern vehicles may limit the maximum allowed speed, increase sensitivity to emergency braking, activate dedicated safety systems, increase the assistance provided by (partially) autonomous systems, or even force a full standstill for safety~\cite{ncap_oms_safety, gsr2019}. A drunk driving detection could also be useful for existing digital intervention solutions outside of the vehicle, such as \emph{Drink Less}, \emph{Daybreak}, or \emph{SoberDiary}~\cite{drinkless, daybreak, SoberDiary2015}. Once informed by our system, such digital interventions could react on the detected drunk driving events and foster alcohol behavior change beyond the driving context.

\subsection{Roadmap to implementation}
\label{sec:outlook}
Our drunk driving detection system can be easily integrated into existing camera-based systems for monitoring driving states such as drowsiness and distraction. These camera systems have the same input data, such as gaze positions, that our system requires, and therefore our system could be directly added as a simple software component (e.g., in~\cite{bosch_dms2022, smarteye2022}).

To bring our system to market, we recommend a few calibration and evaluation steps before our drunk driving detection system is used in real vehicles. First, the current simulator study should be replicated with sober and drunk drivers in a real vehicle. Since driving under the influence of alcohol is a criminal offense in almost all countries (even in the context of a study), the most likeliest way would be to conduct such a study on a test track with a driving instructor next to the driver. Second, our system should prove its ability in everyday traffic. Therefore, we recommend collecting everyday driving data of non-impaired drivers on open roads in a vehicle equipped with a driver monitoring system. This represents a comparably small effort to a drunk driving study as less ethical concerns exist and no clinical trial is needed. The drunk driving study would allow to validate the detection performance of our system in real vehicles, while the open-road study can be used to calibrate against overly sensitive warnings (i.e., false alarms) that are a key threat for long-term adoption and might occur due to the increased diversity of driving conditions in a natural environment.

We also see clear advantages in embedding our drunk driving detection into an existing driver monitoring system on top of existing algorithms for detecting drowsiness or distraction. For example, our drunk driving detection algorithm is validated on the basis of rather undisturbed and focused driving. A distraction detection algorithm can filter out distracted driving to prevent potential erroneous classifications of our detection algorithm. Moreover, having an ensemble of driver monitoring systems for various impairments may further add to the reliability of the overall system.

%% file: body/06_conclusion.tex
\section{Conclusion}

To the best of our knowledge, our system is the first to detect drunk driving from camera-based sensor technology. Thereby, we directly address needs in practice for the \ac{HCI} community: Policy initiatives and regulations around the world (such as in the \US \cite{infrastructurebill2021}) call for new drunk driving detection technologies and interventions and thereby reduce alcohol-related harm. Even though progress has been made toward fully autonomous driving, experts agree that autonomous driving will not be widely available in the next two decades~\cite{autonomousadaption2017, autonomous2018}. Hence, for the coming years, detection systems are needed that build upon existing \ac{HCI} technologies in vehicles that leverage driver vehicle and environment interaction. Here, a cost-effective and scalable approach is offered by our novel machine learning system that uses existing driver monitoring cameras. To this end, our system provides new opportunities for digital interventions to reduce alcohol-related harms, particularly traffic fatalities.

%% file: body/98_appendix.tex
\appendix
\onecolumn
\section{Details on participants}
\label{app:details_participants}
\captionsetup[figure]{name=Supplementary Figure}
\captionsetup[table]{name=Supplementary Table}
\setcounter{figure}{0}
\setcounter{table}{0}

The study flow diagram is shown in \afref{sup:fig:study_flow}. Inclusion and exclusion criteria for participation are explained in the following.

\subsection{Inclusion and exclusion criteria}

\emph{Inclusion criteria} for individuals eligible for participation in the DRIVE study are: (1)~passing the driver examination at least 2 years before study inclusion; (2)~possession of a driver's license that is valid in the European Union or Switzerland; and (3)~reporting moderate alcohol consumption (i.e., neither total absence nor excess). The latter was examined via the \ac{AUDIT}~\citep{audit1993} and the \ac{PEth} level of a capillary blood sample (below 210\,ng/mL)~\citep{wurst2015}.

\emph{Exclusion criteria} for participation in the DRIVE study were if participants met one or more of the following: women who are pregnant, breast feeding, or intend to become pregnant during the course of the study; other clinically significant concomitant disease states as judged by the investigator (e.g., renal failure, hepatic dysfunction, cardiovascular disease etc.); known or suspected non-compliance or drug abuse; inability to follow the procedures of the study, e.g., due to language problems, psychological disorders, dementia, etc.; participation in another study with investigational drug within the 30 days preceding and during the present study; specific concomitant therapy washout requirements prior to and/or during study participation; previous enrollment into the current study; personal dependences with the study team (e.g., employees, family members, and other dependent persons); physical or psychological disease likely to interfere with the normal conduct of the study and interpretation of the study results as judged by the investigator (especially coronary heart disease or epilepsy); current treatment with drugs known to interfere with metabolism (e.g., systemic corticosteroids, statins etc.) or driving performance (e.g., opioids, benzodiazepines); patients not capable of driving with a driving simulator or patients experiencing motion sickness during the simulator test driving session at the introductory study visit.

\subsection{{Study flow}}
\begin{figure*}[h!]
    \centering
    \includegraphics[width=0.7\textwidth]{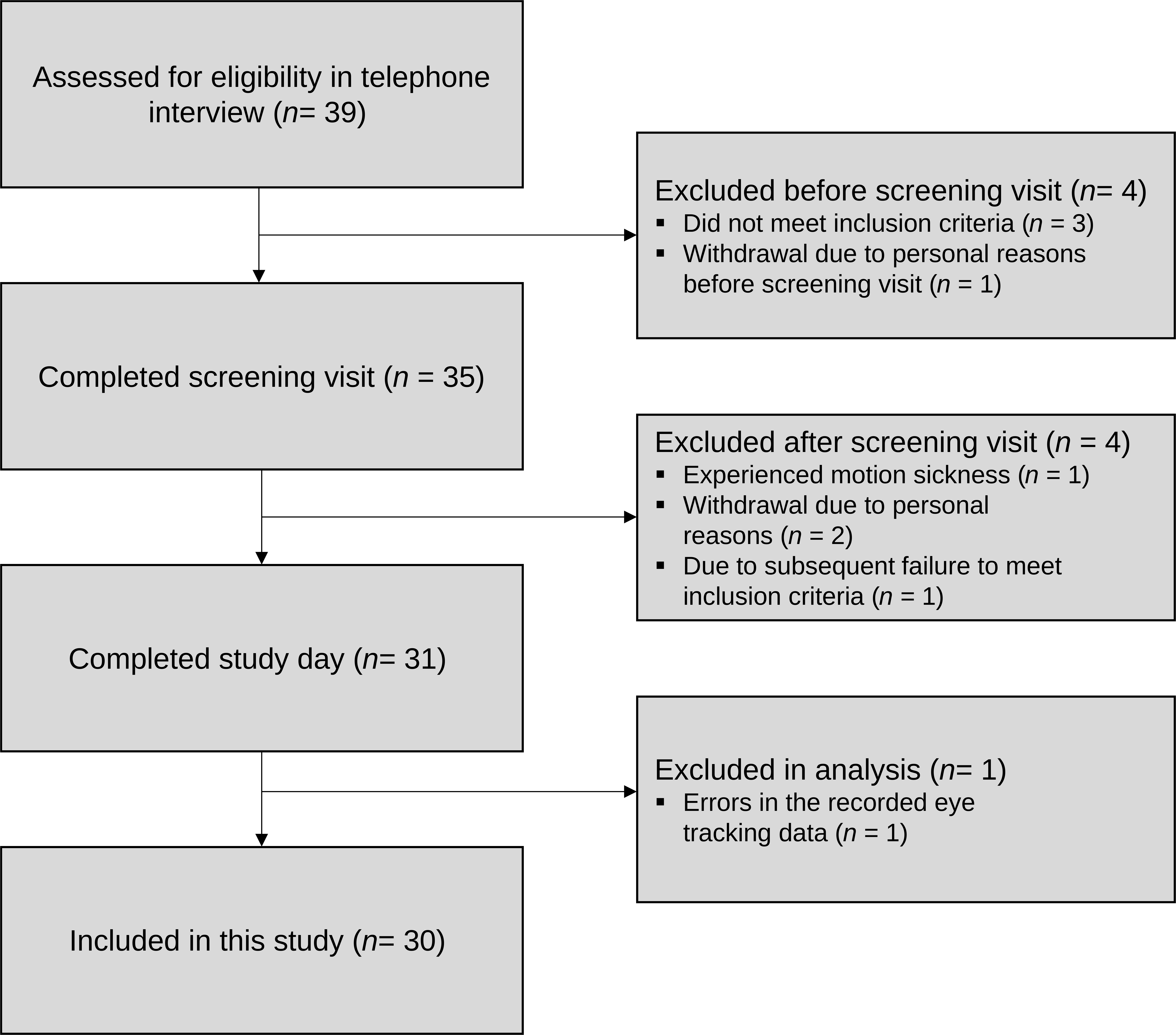}
	\Description{This figure shows a study flow diagram and, thus, the inclusion/exclusion process of the participants in the clinical study. Assessed for eligibility in a telephone interview were 39 participants. Of these, 4 were excluded. 3 did not meet the inclusion criteria and 1 person withdraw before the screening visit due to personal reasons. 35 participants completed the screening visit. Excluded after the screening visit were 4 participants. 1 person experienced motion sickness. 2 withdrew due to personal reasons. 1 person was excluded due to subsequent failure to meet inclusion criteria. 31 participants completed the study day. 1 was in the analysis included due to errors in the recorded eye tracking data. Therefore, 30 participant were in the analysis of this study included.}
 \caption{\textbf{Study flow diagram.} Participants subject to screening and inclusion/exclusion in the clinical study.}~\label{sup:fig:study_flow}
\end{figure*}

Of the screened $39$ participants, we had to exclude $9$ participants as they did not meet the inclusion/exclusion criteria. Three participants were excluded because they did not meet the inclusion criteria from above and reported -- at least occasional -- consumption of illegal drugs. One participant withdrew from the study due to time conflicts. After screening, one individual was excluded due to experiencing motion sickness while driving in the simulator, two participants withdrew for personal reasons, and one because of subsequent failure to comply with the participation requirements (i.e., drug prescription after initial screening). Finally, one individual was excluded due to errors in the eye-tracking recording. A diagram showing the study flow is in~\afref{sup:fig:study_flow}.

\FloatBarrier
\clearpage

\section{Feature generation}
\label{app:feature-generation}

\begin{figure}[h!]
    \centering
    \includegraphics[width=0.8\textwidth]{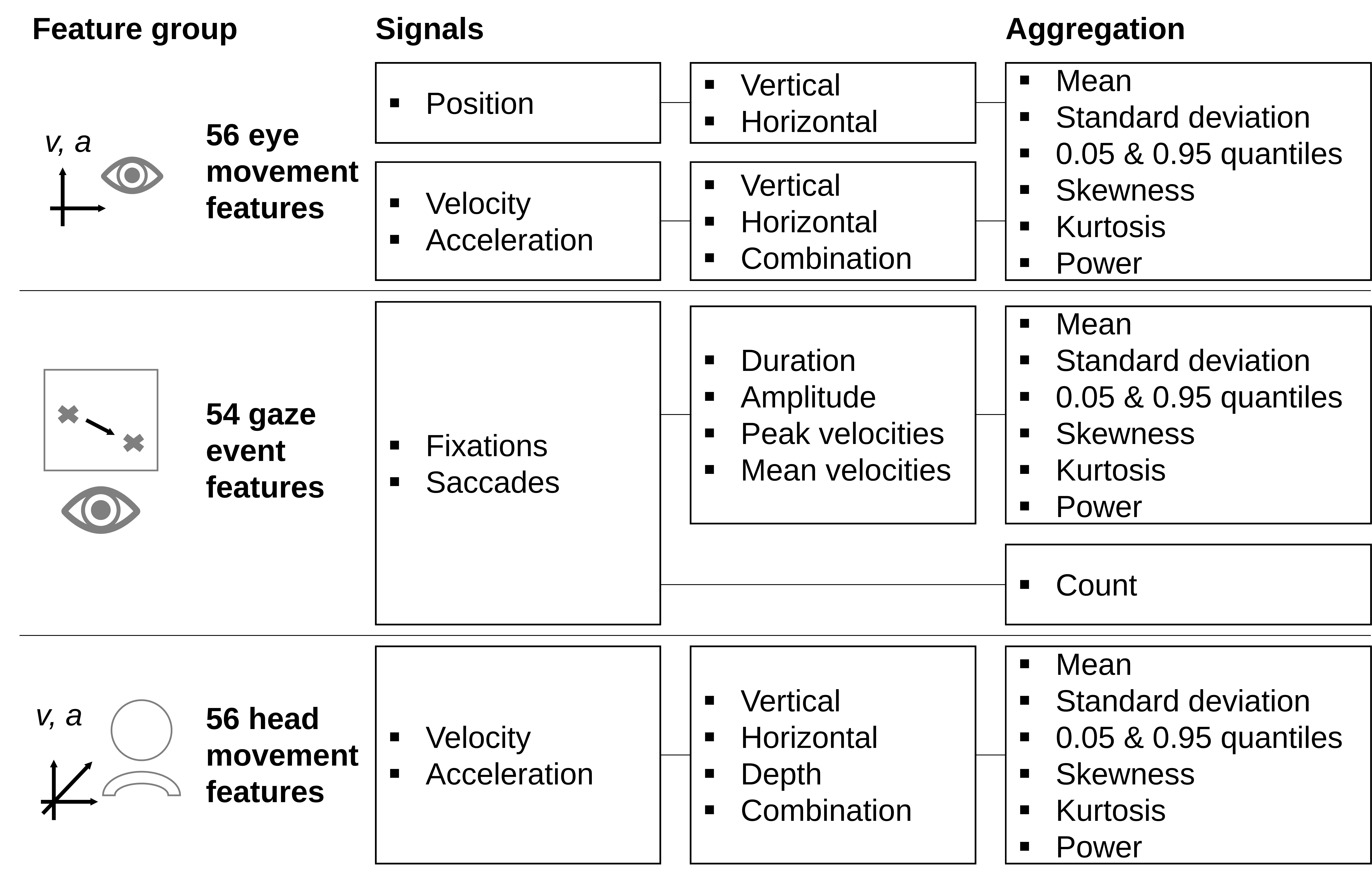}
	\Description{This figure shows an overview of the feature generation. Features are arranged in three groups: eye movement features, gaze event features, and head movements, which are all derived from the driver monitoring camera. In total, we calculated 56 eye movement features, 54 gaze event features, and 56 head movements for each 60 second long window. To calculate these features, we use different signals. For eye movement, our system uses the vertical and horizontal position as well as the vertical, horizontal, and combined velocity and acceleration. For gaze event features, our system uses the fixations and saccades and their respective duration, amplitude as well as the peak and mean velocities. For head movements, our system uses the vertical, horizontal, depth, and their combination of the velocity and acceleration. These signals are further processed by aggregation functions (i.e., to map a time series onto a single value). The aggregation functions are: mean, standard deviation, 0.05-quantile, 0.95-quantile, skewness, kurtosis, and power. In addition, our system takes for fixations and saccades their respective counts.}
 \caption{\textbf{Overview of feature generation.} Features are arranged in three groups: eye movement features, gaze event features, and head movements, which are all derived from the driver monitoring camera. Each feature group is further processed by aggregation functions (i.e., to map a time series onto a single value).}~\label{sup:fig:generated-features}
\end{figure}

The feature generation is shown in \afref{sup:fig:generated-features}.

\FloatBarrier
\clearpage

\section{Additional insights on machine learning performance}
\label{app:additional-results}

\begin{figure}[h!]
    \centering
    \includegraphics[width=0.8\textwidth]{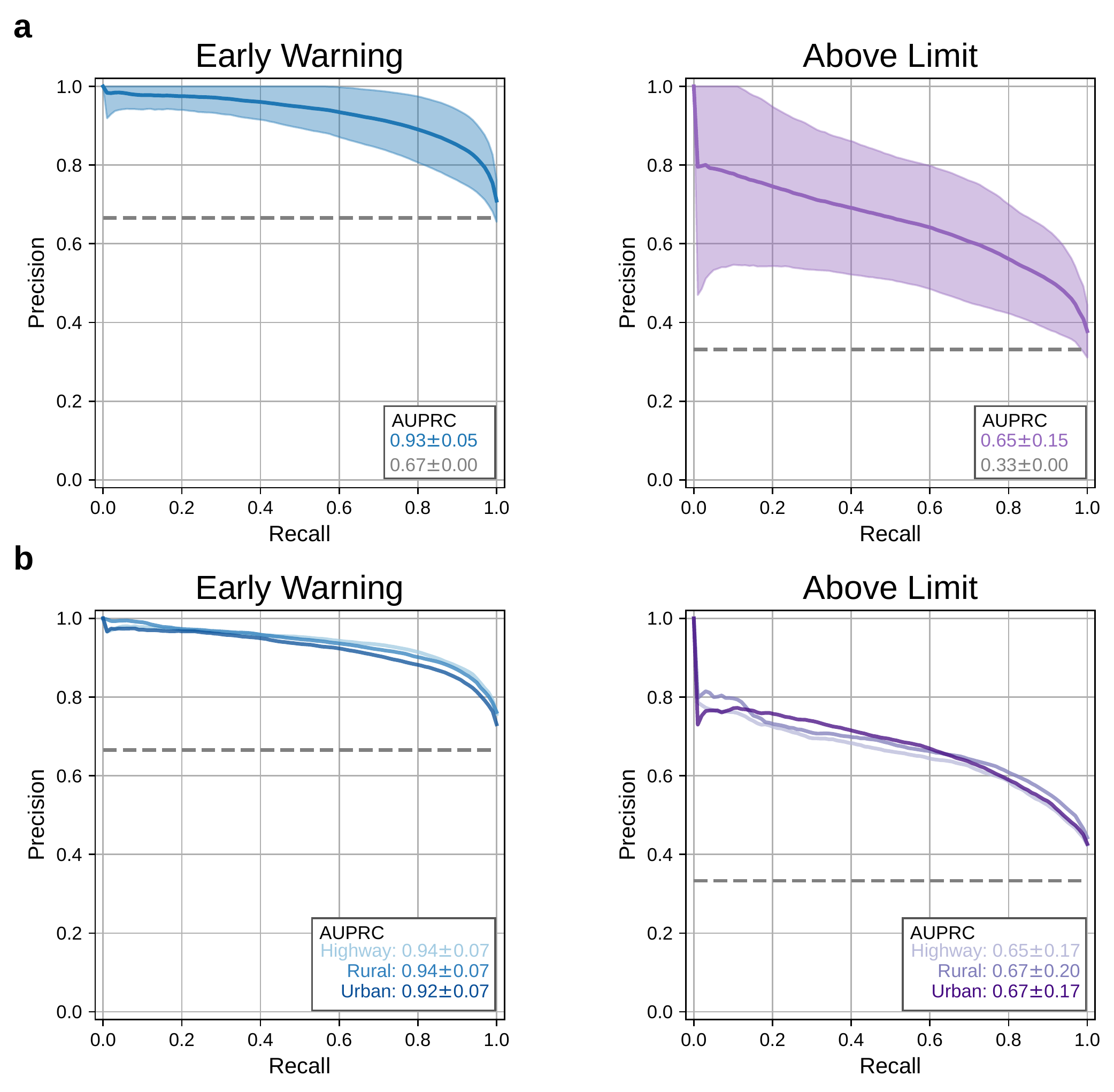}
	\Description{The figure shows AUPRC graphs for EARLY WARNING (left) and ABOVE LIMIT (right) machine learning systems performance. The two graphs in a) (top) show the overall performance. For the EARLY WARNING model, the overall performance is 0.93 ± 0.05 compared to a na{\"i}ve classifier predicting the majority class with 0.67 ± 0.00. For the ABOVE LIMIT model, the overall performance is 0.65 ± 0.05 compared to a na{\"i}ve classifier predicting the majority class with 0.33 ± 0.00. The two graphs in b (bottom) show the performance per scenario. For the EARLY WARNING model, the performance for highway is 0.94 ± .07, for rural is 0.94 ± 0.07, for urban is 0.92 ± 0.07. For the ABOVE LIMIT model, the performance for highway is 0.65 ± 0.17, for rural is 0.67 ± 0.20, for urban is 0.67 ± 0.17.}
 \caption{\textbf{Performance of drunk driving detection.} The machine learning system for detecting drunk driving is evaluated based on the \acf{AUPRC}. \textbf{(a)}~Performance across participants for different BAC thresholds. \textbf{(b)}~Performance by driving scenario (i.e., highway, rural, and urban). The dashed, gray line shows an AUPRC of a na{\"i}ve classifier predicting the majority class, which is consistently outperformed by our \ac{ML} system. \acrexpl{BAC}.}~\label{sup:fig:prc}
\end{figure}

\begin{figure}[h!]
    \centering
    \includegraphics[width=0.85\textwidth]{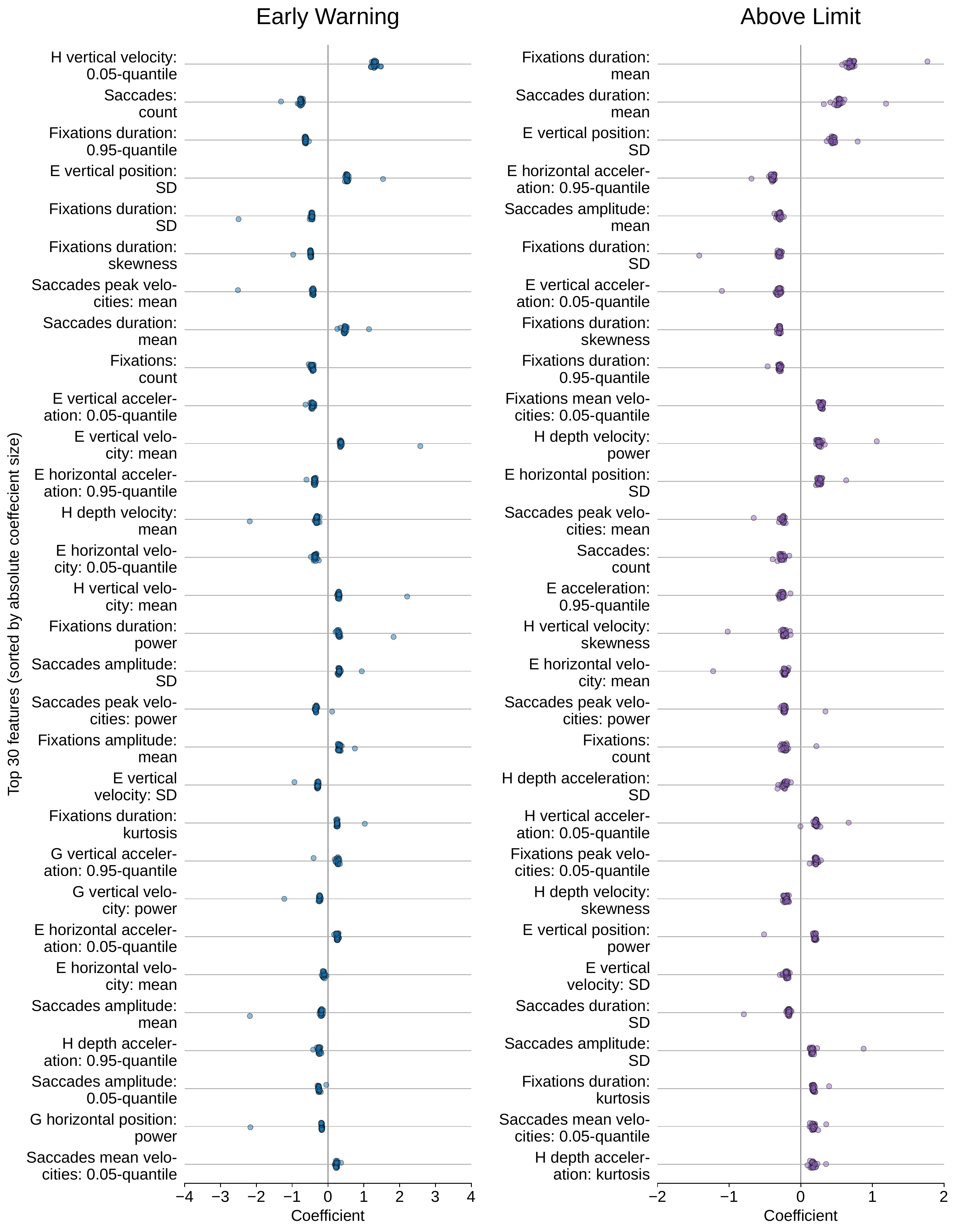}
	\Description{The figure shows the selected features for EARLY WARNING (left) and ABOVE LIMIT (right). Shown are the top 30 features. Features are ranked descending by their absolute coefficient size and, therefore, reflect their importance for the detection task.}
 \caption{\textbf{Selected features in the machine learning system.} Shown are the top 30 features. Features are ranked descending by their absolute coefficient size and, therefore, reflect their importance for the detection task. \acrexpl{E}; \acrexpl{H}; \acrexpl{SD}.}~\label{sup:fig:selected_features}
\end{figure}

Here, we report additional performance metrics (\atref{sup:tab:comparison}, \afref{sup:fig:prc}, and \atref{sup:tab:majority-vote}) and results for machine learning interpretability (\afref{sup:fig:selected_features}).

\begin{table}[!th]
  \centering
  \caption{\textbf{Comparison of performance metrics with additional data sources as a baseline.} We report the performance of our machine learning system while using different data sources: camera-only, CAN-only, and both combined. The following performance metrics are computed: \acf{AUROC}, \acf{AUPRC}, balanced accuracy, and F1 score (weighted by classes). Reported: \meanstd{mean}{\acl{SD}}. \acrexpl{CAN} (i.e., vehicle signals).}
    \begin{tabular}{lcllll}
    \toprule
          
         \textbf{Data}  &  &     \multirow{2}{*}{\textbf{AUROC}} & \multirow{2}{*}{\textbf{AUPRC}} & \textbf{Balanced} & \multirow{2}{*}{\textbf{F1 score}}\\
         \textbf{source} & &       &  & \textbf{accuracy} & \\
    \midrule

     \multicolumn{1}{l}{Camera-} & \cellcolor{early}Early Warning & 0.88$\pm$0.09 & 0.93$\pm$0.05 & 0.76$\pm$0.10 & 0.75$\pm$0.14 \\
    \multicolumn{1}{l}{only} & \cellcolor{above}Above Limit & 0.79$\pm$0.10 & 0.65$\pm$0.16 & 0.68$\pm$0.10 & 0.67$\pm$0.12 \\ \\
    \multicolumn{1}{l}{CAN-} & \cellcolor{early}Early Warning & 0.74$\pm$0.10 & 0.84$\pm$0.07 & 0.65$\pm$0.08 & 0.64$\pm$0.11 \\
    \multicolumn{1}{l}{only} & \cellcolor{above}Above Limit & 0.66$\pm$0.12 & 0.50$\pm$0.15 & 0.60$\pm$0.09 & 0.60$\pm$0.08 \\ \\
    
    \multicolumn{1}{l}{Both} & \cellcolor{early}Early Warning & 0.91$\pm$0.07 & 0.95$\pm$0.04 & 0.78$\pm$0.10 & 0.77$\pm$0.13 \\
    \multicolumn{1}{l}{combined} & \cellcolor{above}Above Limit & 0.81$\pm$0.11 & 0.68$\pm$0.16 & 0.69$\pm$0.10 & 0.69$\pm$0.12 \\

\bottomrule
    \end{tabular}%
  \label{sup:tab:comparison}%

\end{table}%

\clearpage

\begin{table}[!th]
  \centering
  \caption{\textbf{Detailed performance metrics for a majority vote aggregation.} We report the performance of our machine learning system when aggregating non-overlapping single windows over different amount of windows. With an increasing number of of aggregated windows, the performance metrics improve. The following performance metrics are computed: \acf{AUROC}, \acf{AUPRC}, balanced accuracy, and F1 score (weighted by classes). Reported: \meanstd{mean}{\acl{SD}}.}
    \begin{tabular}{lcllll}
    \toprule
          
         \textbf{Window}  &  &     \multirow{2}{*}{\textbf{AUROC}} & \multirow{2}{*}{\textbf{AUPRC}} & \textbf{Balanced} & \multirow{2}{*}{\textbf{F1 score}}\\
         \textbf{amount} & &       &  & \textbf{accuracy} & \\
    \midrule

     \multicolumn{1}{l}{None} & \cellcolor{early}Early Warning & 0.88$\pm$0.09 & 0.93$\pm$0.05 & 0.76$\pm$0.10 & 0.75$\pm$0.14 \\
    & \cellcolor{above}Above Limit & 0.79$\pm$0.10 & 0.65$\pm$0.16 & 0.68$\pm$0.10 & 0.67$\pm$0.12 \\ \\
    
    \multicolumn{1}{l}{5} & \cellcolor{early}Early Warning & 0.88$\pm$0.08 & 0.93$\pm$0.05 & 0.76$\pm$0.10 & 0.75$\pm$0.13 \\
    & \cellcolor{above}Above Limit & 0.80$\pm$0.11 & 0.66$\pm$0.16 & 0.68$\pm$0.09 & 0.67$\pm$0.12 \\ \\
    
    \multicolumn{1}{l}{30} & \cellcolor{early}Early Warning & 0.89$\pm$0.08 & 0.93$\pm$0.05 & 0.75$\pm$0.11 & 0.75$\pm$0.13 \\
    & \cellcolor{above}Above Limit & 0.80$\pm$0.10 & 0.67$\pm$0.15 & 0.67$\pm$0.10 & 0.67$\pm$0.12 \\ \\
    
    \multicolumn{1}{l}{60} & \cellcolor{early}Early Warning & 0.90$\pm$0.08 & 0.94$\pm$0.04 & 0.77$\pm$0.11 & 0.76$\pm$0.13 \\
    & \cellcolor{above}Above Limit & 0.82$\pm$0.11 & 0.70$\pm$0.15 & 0.68$\pm$0.11 & 0.67$\pm$0.14 \\ \\
    
    \multicolumn{1}{l}{90} & \cellcolor{early}Early Warning & 0.90$\pm$0.08 & 0.93$\pm$0.05 & 0.76$\pm$0.11 & 0.75$\pm$0.13 \\
    & \cellcolor{above}Above Limit & 0.83$\pm$0.11 & 0.71$\pm$0.15 & 0.68$\pm$0.11 & 0.67$\pm$0.13 \\ \\

    \multicolumn{1}{l}{120} & \cellcolor{early}Early Warning & 0.91$\pm$0.08 & 0.94$\pm$0.06 & 0.78$\pm$0.12 & 0.77$\pm$0.14 \\
    & \cellcolor{above}Above Limit & 0.84$\pm$0.11 & 0.75$\pm$0.16 & 0.68$\pm$0.12 & 0.67$\pm$0.15 \\ \\

    \multicolumn{1}{l}{150} & \cellcolor{early}Early Warning & 0.91$\pm$0.08 & 0.94$\pm$0.06 & 0.79$\pm$0.13 & 0.77$\pm$0.15 \\
    & \cellcolor{above}Above Limit & 0.85$\pm$0.11 & 0.76$\pm$0.17 & 0.68$\pm$0.13 & 0.67$\pm$0.16 \\ \\

    \multicolumn{1}{l}{180} & \cellcolor{early}Early Warning & 0.88$\pm$0.09 & 0.89$\pm$0.09 & 0.73$\pm$0.12 & 0.72$\pm$0.15 \\
    & \cellcolor{above}Above Limit & 0.83$\pm$0.10 & 0.71$\pm$0.14 & 0.66$\pm$0.12 & 0.65$\pm$0.15 \\

\bottomrule
    \end{tabular}%
  \label{sup:tab:majority-vote}%
\end{table}%

\clearpage

\FloatBarrier

\section{Robustness checks}
\label{app:robustness-checks}

\begin{figure}[h!]
    \centering
    \includegraphics[width=\textwidth]{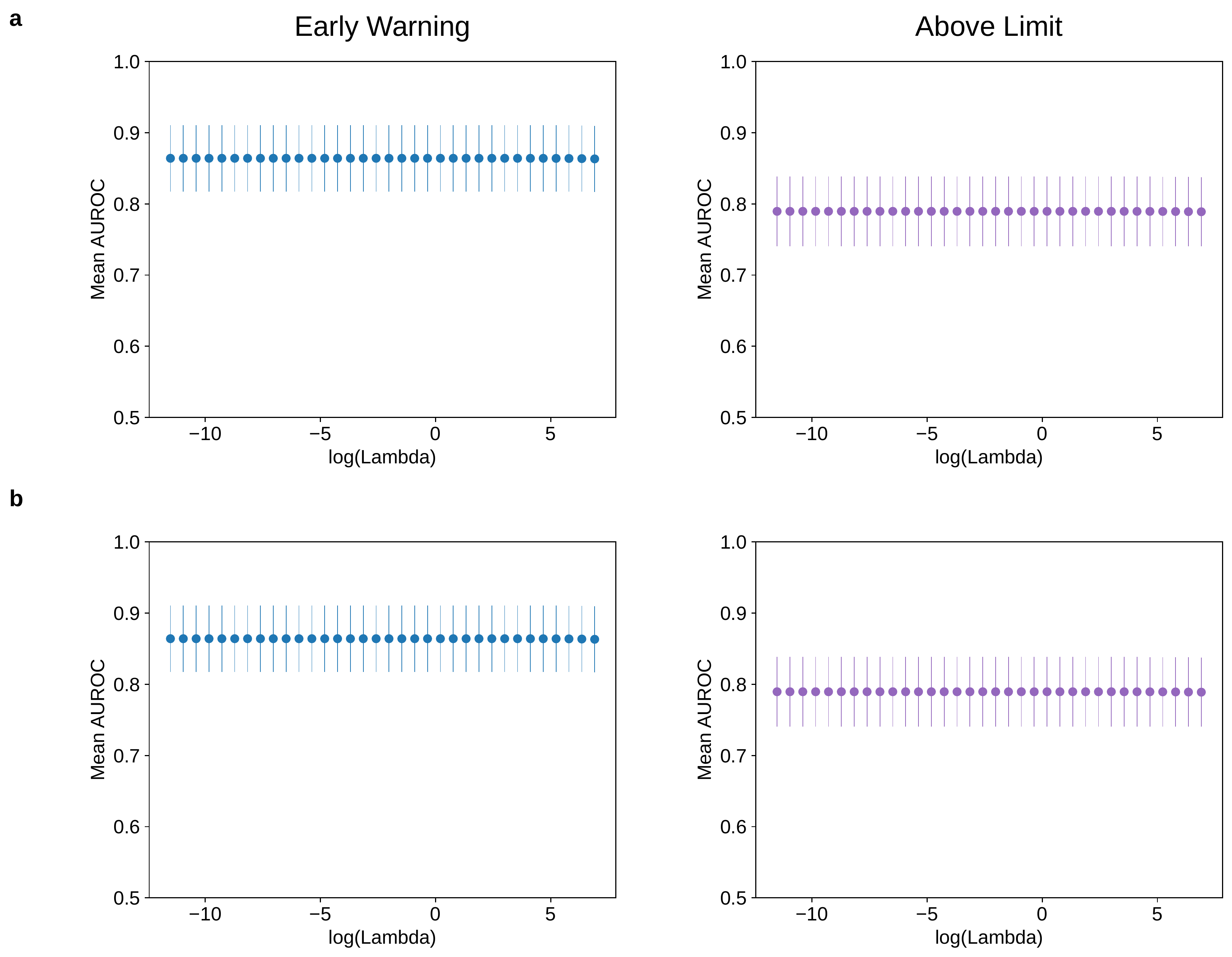}
	\Description{Sensitivity analysis for regularization. The following evaluations compare the performance when varying the hyperparameter lambda inside the regularization. Across all lambdas, the machine learning system yields a robust, high prediction performance. a) The prediction performance for different hyperparameters (lambda) that weight the L1 penalty in a logistic regression with lasso regularization. b) The prediction for different hyperparameters (lambda) that weight the L2 penalty in a logistic regression with ridge regularization. With varying log(Lambda) values from -12 to 6, the performance does not change. For both classifiers, the mean AUROC performance is for EARLY WARNING 0.88 and for ABOVE LIMIT 0.79 for every log(Lambda) value.}
 \caption{\textbf{Sensitivity analysis for regularization.} The following evaluations compare the performance when varying the hyperparameter lambda inside the regularization. Across all lambdas, the machine learning system yields a robust, high prediction performance. \textbf{(a)}~The prediction performance for different hyperparameters (lambda) that weight the L1 penalty in a logistic regression with lasso regularization. \textbf{(b)}~The prediction for different hyperparameters (lambda) that weight the L2 penalty in a logistic regression with ridge regularization. Reported: \meanstd{mean}{\acl{SD}}. \acrexpl{AUROC}.}~\label{sup:fig:lambda}
\end{figure}

Here, we report the following robustness checks for the regularization (\afref{sup:fig:lambda}), an evaluation based on leaving one scenario in training out and evaluating on it (\atref{sup:tab:unseen-scenarios}), the length of the sliding window (\atref{sup:tab:window-analysis}), the predictive power of different feature groups (\atref{sup:tab:feature-analysis}), and alternative \ac{ML} models (\atref{sup:tab:classifier_analysis}). 

\begin{table}[!th]
\caption{\textbf{Sensitivity analysis for leaving-one-driving-scenario-out.} The following evaluations compare the performance when our machine learning system is trained on two driving scenarios and is evaluated on on the third driving scenario. This analysis is done on top of our leave-one-subject out validation. The machine learning system has a robust, high prediction performance even for previously unseen scenarios. Reported: \meanstd{mean}{\acl{SD}}. \acrexpl{AUPRC}; \acrexpl{AUROC}.}~\label{sup:tab:unseen-scenarios}
\centering
    \begin{tabular}{rlllll}
    \toprule
    \multicolumn{1}{l}{\textbf{Unseen}} & \textbf{Task} &  \textbf{AUROC} & \textbf{AUPRC} & \textbf{Balanced} & \textbf{F1 score} \\
    \multicolumn{1}{l}{\textbf{scenario}} &  & & & \textbf{accuracy} & \\
    \midrule
    \multicolumn{1}{l}{Highway} & \cellcolor{early}Early Warning & 0.89$\pm$0.12 & 0.93$\pm$0.08 & 0.75$\pm$0.13 & 0.77$\pm$0.13 \\
          & \cellcolor{above}Above Limit & 0.78$\pm$0.13 & 0.64$\pm$0.17 & 0.66$\pm$0.12 & 0.60$\pm$0.17 \\
          &       &       &       &       &  \\
    \multicolumn{1}{l}{Rural} & \cellcolor{early}Early Warning & 0.94$\pm$0.05 & 0.88$\pm$0.06 & 0.76$\pm$0.14 & 0.75$\pm$0.15 \\
          & \cellcolor{above}Above Limit & 0.77$\pm$0.18 & 0.64$\pm$0.20 & 0.66$\pm$0.13 & 0.64$\pm$0.15 \\
          &       &       &       &       &  \\
    \multicolumn{1}{l}{Urban} & \cellcolor{early}Early Warning & 0.92$\pm$0.07 & 0.91$\pm$0.05 & 0.73$\pm$0.10 & 0.70$\pm$0.17 \\
          & \cellcolor{above}Above Limit & 0.80$\pm$0.11 & 0.65$\pm$0.16 & 0.66$\pm$0.11 & 0.67$\pm$0.11 \\
          &       &       &       &       &  \\
    \multicolumn{1}{l}{Averaged} & \cellcolor{early}Early Warning & 0.82$\pm$0.11 & 0.93$\pm$0.07 & 0.75$\pm$0.12 & 0.74$\pm$0.15 \\
      & \cellcolor{above}Above Limit & 0.78$\pm$0.14 & 0.64$\pm$0.18 & 0.66$\pm$0.12 & 0.64$\pm$0.15 \\
    \bottomrule
    \end{tabular}%
\end{table}

\clearpage

\begin{table}[!th]
\caption{\textbf{Sensitivity analysis for window size.} The following evaluations compare the performance when varying the size of the sliding window. Across all sizes, the machine learning system has a robust, high prediction performance. Reported: \meanstd{mean}{\acl{SD}}. \acrexpl{AUPRC}; \acrexpl{AUROC}.}~\label{sup:tab:window-analysis}
\centering
    \begin{tabular}{rlllll}
    \toprule
    \multicolumn{1}{l}{\textbf{Window}} & \textbf{Task} &  \textbf{AUROC} & \textbf{AUPRC} & \textbf{Balanced} & \textbf{F1 score} \\
    \multicolumn{1}{l}{\textbf{size}} &  & & & \textbf{accuracy} & \\
    \midrule
    \multicolumn{1}{l}{5s} & \cellcolor{early}Early Warning & 0.70$\pm$0.08 & 0.81$\pm$0.06 & 0.63$\pm$0.07 & 0.64$\pm$0.09 \\
          & \cellcolor{above}Above Limit & 0.65$\pm$0.06 & 0.45$\pm$0.07 & 0.60$\pm$0.05 & 0.60$\pm$0.06 \\
          &       &       &       &       &  \\
    \multicolumn{1}{l}{10s} & \cellcolor{early}Early Warning & 0.75$\pm$0.09 & 0.84$\pm$0.06 & 0.67$\pm$0.08 & 0.67$\pm$0.10 \\
          & \cellcolor{above}Above Limit & 0.69$\pm$0.07 & 0.50$\pm$0.09 & 0.63$\pm$0.06 & 0.63$\pm$0.08 \\
          &       &       &       &       &  \\
    \multicolumn{1}{l}{20s} & \cellcolor{early}Early Warning & 0.81$\pm$0.10 & 0.88$\pm$0.06 & 0.71$\pm$0.09 & 0.71$\pm$0.11 \\
          & \cellcolor{above}Above Limit & 0.74$\pm$0.09 & 0.56$\pm$0.11 & 0.65$\pm$0.08 & 0.65$\pm$0.10 \\
          &       &       &       &       &  \\
    \multicolumn{1}{l}{40s} & \cellcolor{early}Early Warning & 0.85$\pm$0.10 & 0.91$\pm$0.05 & 0.74$\pm$0.10 & 0.73$\pm$0.13 \\
          & \cellcolor{above}Above Limit & 0.77$\pm$0.10 & 0.62$\pm$0.14 & 0.67$\pm$0.09 & 0.66$\pm$0.12 \\
          &       &       &       &       &  \\
    \multicolumn{1}{l}{60s} & \cellcolor{early}Early Warning & 0.88$\pm$0.09 & 0.93$\pm$0.05 & 0.76$\pm$0.10 & 0.75$\pm$0.14 \\
          & \cellcolor{above}Above Limit & 0.79$\pm$0.10 & 0.65$\pm$0.16 & 0.68$\pm$0.10 & 0.67$\pm$0.12 \\
          &       &       &       &       &  \\
    \multicolumn{1}{l}{80s} & \cellcolor{early}Early Warning & 0.89$\pm$0.09 & 0.94$\pm$0.04 & 0.77$\pm$0.10 & 0.76$\pm$0.13 \\
          & \cellcolor{above}Above Limit & 0.80$\pm$0.12 & 0.67$\pm$0.17 & 0.68$\pm$0.11 & 0.67$\pm$0.13 \\
          &       &       &       &       &  \\
    \multicolumn{1}{l}{120s} & \cellcolor{early}Early Warning & 0.91$\pm$0.08 & 0.95$\pm$0.04 & 0.79$\pm$0.11 & 0.78$\pm$0.13 \\
          & \cellcolor{above}Above Limit & 0.82$\pm$0.12 & 0.69$\pm$0.19 & 0.68$\pm$0.12 & 0.67$\pm$0.14 \\
    \bottomrule
    \end{tabular}%
\end{table}

\clearpage

\begin{table}[!th]
\caption{\textbf{Sensitivity analysis for feature group.} The following evaluations compare the performance when varying the selected features. Across all the features groups, gaze events perform best. However, the prediction performance drops compared to combining all features. Reported: \meanstd{mean}{\acl{SD}}. \acrexpl{AUPRC}; \acrexpl{AUROC}.}~\label{sup:tab:feature-analysis}
\centering
    \begin{tabular}{llllll}
    \toprule
    \multicolumn{1}{l}{\textbf{Feature}} & \textbf{Task} &   \textbf{AUROC}    &    \textbf{AUPRC}   & \textbf{Balanced} & \textbf{F1 score} \\
    \multicolumn{1}{l}{\textbf{group}} & \textbf{} &  &  & \textbf{accuracy} & \\
    \midrule
    Eye   & \cellcolor{early}Early Warning & 0.79$\pm$0.11 & 0.87$\pm$0.07 & 0.69$\pm$0.10 & 0.69$\pm$0.12 \\
    movements & \cellcolor{above}Above Limit & 0.76$\pm$0.10 & 0.60$\pm$0.14 & 0.67$\pm$0.09 & 0.65$\pm$0.11 \\
          &       &       &       &       &  \\
    Gaze  & \cellcolor{early}Early Warning & 0.84$\pm$0.09 & 0.91$\pm$0.06 & 0.71$\pm$0.11 & 0.70$\pm$0.15 \\
    events & \cellcolor{above}Above Limit & 0.79$\pm$0.08 & 0.63$\pm$0.13 & 0.68$\pm$0.08 & 0.67$\pm$0.11 \\
          &       &       &       &       &  \\
    Head  & \cellcolor{early}Early Warning & 0.82$\pm$0.12 & 0.89$\pm$0.09 & 0.67$\pm$0.11 & 0.65$\pm$0.17 \\
    movements & \cellcolor{above}Above Limit & 0.69$\pm$0.14 & 0.54$\pm$0.16 & 0.62$\pm$0.11 & 0.59$\pm$0.16 \\
              \bottomrule
    \end{tabular}%
\end{table}

\clearpage

\begin{table}[htbp]
  \centering
  \caption{\textbf{Sensitivity analysis for machine learning model.} The following evaluations compare the performance when varying the machine learning model. Across all models, the machine learning system has a robust, high prediction performance. Reported: \meanstd{mean}{\acl{SD}}. \acrexpl{AUPRC}; \acrexpl{AUROC}; \acrexpl{MLP}; \acrexpl{SVM}.}%
  \label{sup:tab:classifier_analysis}%
    \begin{tabular}{rlllll}
    \toprule
         \multicolumn{1}{l}{\textbf{Model}}  & \textbf{Task} &    \textbf{AUROC}   &   \textbf{AUPRC}    & \textbf{Balanced} & \textbf{F1 score}  \\
    & &  & & \textbf{accuracy} &  \\
    \midrule
    \multicolumn{1}{l}{Lasso} & \cellcolor{early}Early Warning & 0.88$\pm$0.09 & 0.93$\pm$0.05 & 0.76$\pm$0.10 & 0.75$\pm$0.14 \\
          & \cellcolor{above}Above Limit & 0.79$\pm$0.10 & 0.65$\pm$0.16 & 0.68$\pm$0.10 & 0.67$\pm$0.12 \\
          &       &       &       &       &  \\
    \multicolumn{1}{l}{Ridge} & \cellcolor{early}Early Warning & 0.88$\pm$0.09 & 0.93$\pm$0.05 & 0.76$\pm$0.11 & 0.75$\pm$0.14 \\
          & \cellcolor{above}Above Limit & 0.79$\pm$0.10 & 0.65$\pm$0.16 & 0.68$\pm$0.10 & 0.67$\pm$0.12 \\
          &       &       &       &       &  \\
    \multicolumn{1}{l}{Elastic Net} & \cellcolor{early}Early Warning & 0.88$\pm$0.09 & 0.93$\pm$0.05 & 0.76$\pm$0.10 & 0.75$\pm$0.14 \\
          & \cellcolor{above}Above Limit & 0.79$\pm$0.10 & 0.65$\pm$0.16 & 0.68$\pm$0.10 & 0.67$\pm$0.12 \\
          &       &       &       &       &  \\
    \multicolumn{1}{l}{SVM} & \cellcolor{early}Early Warning & 0.83$\pm$0.12 & 0.90$\pm$0.08 & 0.71$\pm$0.09 & 0.73$\pm$0.10 \\
    & \cellcolor{above}Above Limit & 0.73$\pm$0.12 & 0.57$\pm$0.15 & 0.61$\pm$0.09 & 0.64$\pm$0.08 \\
      &       &       &       &       &  \\
    \multicolumn{1}{l}{Random} & \cellcolor{early}Early Warning & 0.79$\pm$0.14 & 0.87$\pm$0.09 & 0.61$\pm$0.12 & 0.65$\pm$0.12 \\
    \multicolumn{1}{l}{forest} & \cellcolor{above}Above Limit & 0.73$\pm$0.09 & 0.55$\pm$0.13 & 0.57$\pm$0.08 & 0.61$\pm$0.08 \\
          &       &       &       &       &  \\
    \multicolumn{1}{l}{Gradient} & \cellcolor{early}Early Warning & 0.83$\pm$0.13 & 0.90$\pm$0.08 & 0.68$\pm$0.12 & 0.71$\pm$0.12 \\
    \multicolumn{1}{l}{boosting} & \cellcolor{above}Above Limit & 0.77$\pm$0.08 & 0.60$\pm$0.11 & 0.61$\pm$0.08 & 0.65$\pm$0.08 \\
          &       &       &       &       &  \\
    \multicolumn{1}{l}{MLP} & \cellcolor{early}Early Warning & 0.85$\pm$0.10 & 0.92$\pm$0.06 & 0.73$\pm$0.09 & 0.75$\pm$0.09 \\
          & \cellcolor{above}Above Limit & 0.73$\pm$0.12 & 0.56$\pm$0.15 & 0.61$\pm$0.09 & 0.64$\pm$0.08 \\
              \bottomrule
    \end{tabular}
\end{table}%